\input harvmac
\input epsf


\def\figin{\epsfcheck\figin}\def\figins{\epsfcheck\figins}
\def\epsfcheck{\ifx\epsfbox\UnDeFiNeD
\message{(NO epsf.tex, FIGURES WILL BE IGNORED)}
\gdef\figin##1{\vskip2in}\gdef\figins##1{\hskip.5in}
\else\message{(FIGURES WILL BE INCLUDED)}%
\gdef\figin##1{##1}\gdef\figins##1{##1}\fi}
\def\DefWarn#1{}
\def\figinsert{\goodbreak\topinsert}
\def\ifig#1#2#3#4{\DefWarn#1\xdef#1{fig.~\the\figno}
\writedef{#1\leftbracket fig.\noexpand~\the\figno}%
\figinsert\figin{\centerline{\epsfxsize=#3mm \epsfbox{#2}}}
\bigskip\medskip\centerline{\vbox{\baselineskip12pt
\advance\hsize by -1truein\noindent\footnotefont{\sl Fig.~\the\figno:}\sl\ #4}}
\bigskip\endinsert\noindent\global\advance\figno by1}

\def\ndt{{\noindent}}

\lref\whee{J. A. Wheeler, in ``Relativity, Groups and Topology'', edited
by B. S. and C. M. DeWitt (Gordon and Breach, New York, 1964).}

\lref\hawk{S. W. Hawking, ``Space-time Foam'', Nucl. Phys. {\bf B144} (1978)
349-362.}

\lref\wittos{E. Witten, ``Two-dimensional gravity and intersection theory on moduli space'',
Surveys Diff. Geom. 1 (1991) 243-310.}

\lref\besa{M. Bershadsky, V. Sadov, ``Theory of K\"ahler Gravity'', Int. J. Mod. Phys. {\bf A11} (1996) 4689-4730, {\tt hep-th/9410011}.}

\lref\gova{R. Gopakumar, C. Vafa, ``On the Gauge Theory/Geometry
Correspondence'', Adv. Theor. Math. Phys. 3 (1999) 1415-1443, {\tt hep-th/9811131}}.

\lref\bov{N. Berkovits, H. Ooguri, C. Vafa, ``On the Worldsheet Derivation
of Large N Dualities for the Superstring'', {\tt hep-th/0310118}.}

\lref\bcov{M. Bershadsky, S. Cecotti, H. Ooguri, C. Vafa, ``Kodaira-Spencer
Theory of Gravity and Exact Results for Quantum String Amplitudes'', Commun. Math. Phys. {\bf 165} (1994) 311-428, {\tt hep-th/9309140}.}

\lref\naret{I. Antoniadis, E. Gava, K. S. Narain, T. R. Taylor, ``Topological
Amplitudes in String Theory'', Nucl. Phys. {\bf B413} (1994) 162-184, {\tt hep-th/9307158}.}

\lref\orv{A. Okounkov, N. Reshetikhin, C. Vafa, ``Quantum Calabi-Yau and
Classical Crystals'', {\tt hep-th/0309208}.}

\lref\osv{H. Ooguri, A. Strominger, C. Vafa, work in progress.}

\lref\witcs{E. Witten, ``Chern-Simons Gauge Theory as a String Theory'',
Prog. Math. {\bf 133} (1995) 637-678, {\tt hep-th/9207094}.}

\lref\tvtwo{M. Aganagic, A. Klemm, M. Marino, C. Vafa,
``The topological vertex'',
{\tt hep-th/0305132}.}

\lref\dijky{R. Dijkgraaf, C. Vafa, ``N=1 supersymmetry,
deconstruction, and bosonic gauge theories,'' {\tt hep-th/0302011}.}

\lref\warny{M. Aganagic, K. Intriligator, C. Vafa, N.P. Warner,
``The Glueball Superpotential,'' {\tt hep-th/0304271}.}

\lref\nekin{N. Nekrasov, ``Noncommutative Instantons Revisited",
Comm. Math. Phys. {\bf 241} (2003) 143-160, {\tt hep-th/0010017}.}

\lref\nekb{H. W. Braden, N. Nekrasov, ``Space-Time Foam From
Non-Commutative Instantons'', {\tt hep-th/9912019}.}

\lref\guko{
S. Gukov, ``Three-dimensional quantum gravity, Chern-Simons
theory and the A-polynomial,'' {\tt hep-th/0306165}.}

\lref\wittencsg{E. Witten, ``Topology changing amplitudes in (2+1)-dimensional gravity'', Nucl. Phys. {\bf B323} 113, 1989.}

\lref\lins{E. Witten, ``Phases of N=2 theories in two dimensions'',  Nucl. Phys. {\bf B403} (1993) 159-222, {\tt hep-th/9301042}.}

\lref\mic{M. Aganagic, C. Vafa, ``Mirror symmetry, D-branes and counting
holomorphic disks'', {\tt hep-th/0012041}.}

\lref\guv{S. Gukov, C. Vafa, unpublished.}

\lref\hv{K. Hori, C. Vafa, ``Mirror symmetry'', {\tt hep-th/0002222}.}

\lref\hiv{K. Hori, A. Iqbal, C. Vafa, ``Mirror symmetry and D-branes'',
{\tt hep-th/0005247}.}

\lref\ov{H. Ooguri, C. Vafa, ``Knot invariants and topological strings'',
Nucl. Phys. {\bf B577} (2000) 419-438, {\tt hep-th/991213}.}

\lref\hh{A. Hanany, K. Hori, "Branes and $N=2$ theories in two dimensions", Nucl. Phys. {\bf B513} (1998) 119-174, {\tt hep-th/9707192}.}

\lref\hv{K. Hori, C. Vafa, "Mirror symmetry", {\tt hep-th/0002222}.}

\lref\tvtwoo{M. Aganagic, R. Dijkgraaf, A. Klemm, M. Marino, C. Vafa, ``Topological
strings and integrable hierarchies,'' to appear.}

\lref\cym{R. Cerf, R. Kenyon, ``The low-temperature expansion of the Wulff
crystal in the 3D Ising model'', Comm. Math. Phys. {\bf 222} (2001), no. 1,
147-179\semi
A. Okounkov, N. Reshetikhin, Journal of the American Mathematical Society, {\bf 16}, no. 3 (2003), pp. 581-603.}

\lref\ihiggs{G.~Moore, N.~Nekrasov,
S.~Shatashvili, ``Integrating over the Higgs branches'', {\tt hep-th/9712241}\semi
G.~Moore, N.~Nekrasov,
S.~Shatashvili, ``D-particle bound states and generalized instantons'',
 {\tt hep-th/9803265}. }

\lref\bks{
L.~Baulieu, H.~Kanno, and I.~Singer, ``Special quantum field theories
in eight and other dimensions'', {\tt hep-th/9704167}\semi
B.~Acharya, M.~O'Loughlin, B.~Spence, ``Higher dimensional analogues
of Donaldson-Witten theory'', {\tt hep-th/9705138}\semi
M.~Blau, G.~Thompson, ``Euclidean SYM theories by time reduction
and special holonomy manifolds,'' Phys. Lett. {\bf B415} (1997) 242-252,
{\tt hep-th/9706225}.}

\lref\thom{S.~Donaldson and R.~Thomas, ``Gauge theory
in higher dimensions'', in
{\it The geometric universe: science, geometry, and the work of Roger Penrose}, S. Huggett et.
al eds., Oxford Univ. Press, 1998.}

\lref\kraus{P.~Kraus, M.~Shigemori, ``Noncommutative instantons and the Seiberg-Witten
map,'' {\tt hep-th/0110035}.}

\lref\neksw{N.~Nekrasov, ``Seiberg-Witten prepotential via instanton
counting'', {\tt hep-th/0206161}.}

\lref\no{N.~Nekrasov, A.~Okounkov, ``Seiberg-Witten theory and random
partitions'', {\tt hep-th/0306238}.}

\lref\nakajima{H.~Nakajima, ``Lectures on Hilbert Schemes of
Points on Surfaces''\semi AMS University Lecture Series, 1999,
ISBN 0-8218-1956-9. }

\lref\mnop{D. ~Maulik, N. ~Nekrasov, A. ~Okounkov, R.
~Pandharipande, "Gromov-Witten theory and Donaldson-Thomas
theory", {\tt math.AG/0312059}.}

\lref\bsftop{M.~Bershadsky, V.~Sadov, C.~Vafa, ``D-branes and topological
field theories,''
Nucl.\ Phys.\  {\bf B463}, (1996) 420, {\tt hep-th/9511222}.}

\lref\neksch{N.~Nekrasov, A.~S.~Schwarz, ``Instantons on noncommutative $R^4$ and $(2,0)$
superconformal six-dimensional theory,''
 \cmp{198}{1998}{689}, {\tt hep-th/9802068}.}

\lref\nikloc{N.~Nekrasov, ``Localizing gauge theory'', lecture notes, to appear.}

\lref\ny{H.~Nakajima, K.~Yoshioka, "Instanton counting on blowup, I", {\tt math.AG/0306198}.}

\lref\opennc{N.~Nekrasov, ``Lectures on open strings'', {\tt hep-th/0203109}.}

\lref\donsw{E.~Witten, ``Monopoles and Four-Manifolds'',
Math.~Res.~Lett. {\bf 1} (1994) 769-796, hep-th/9411102}

\lref\nfour{C.~Vafa, E.~Witten, ``A Strong Coupling Test of S-Duality'',
Nucl.Phys. B431 (1994) 3-77, hep-th/9408074}

\lref\DHf{J.~J.~Duistermaat, G.J.~Heckman, Invent.
Math. {\bf 69} (1982) 259\semi M.~Atiyah, R.~Bott, Topology {\bf
23} No 1 (1984) 1}

\lref\witsei{N.~Seiberg, E.~Witten, ``String theory and noncommutative
geometry,''
\jhep{9909}{1999}{032}, {\tt , hep-th/9908142}.}

\lref\loopqg{T. Thiemann, ``Lectures on loop quantum gravity,''
Lect. Notes Phys. {\bf 631} (2003) 41-135, {\tt gr-qc/0210094}.}


\def\boxit#1{\vbox{\hrule\hbox{\vrule\kern8pt
\vbox{\hbox{\kern8pt}\hbox{\vbox{#1}}\hbox{\kern8pt}}
\kern8pt\vrule}\hrule}}
\def\mathboxit#1{\vbox{\hrule\hbox{\vrule\kern8pt\vbox{\kern8pt
\hbox{$\displaystyle #1$}\kern8pt}\kern8pt\vrule}\hrule}}

\def\cmp#1#2#3{Comm. Math. Phys. {\bf #1} (#2) #3}

\def\jhep#1#2#3{JHEP {\bf#1}(#2) #3}

\def\a{{\alpha}}

\def\b{{\beta}}
\def\d{{\delta}}

\def\g{{\gamma}}
\def\e{{\epsilon}}

\def\ve{{\varepsilon}}

\def\m{{\mu}}
\def\n{{\nu}}
\def\u{{\Upsilon}}
\def\l{{\lambda}}

\def\t{{\theta}}



\chardef\tempcat=\the\catcode`\@ \catcode`\@=11
\def\cyracc{\def\u##1{\if \i##1\accent"24 i%
    \else \accent"24 ##1\fi }}
\newfam\cyrfam
\font\tencyr=wncyr10
\def\cyr{\fam\cyrfam\tencyr\cyracc}


\def\CE{{\cal E}}

\def\CH{{\cal H}}
\def\CI{{\cal I}}

\def\CL{{\cal L}}
\def\CM{{\cal M}}
\def\CN{{\cal N}}
\def\CO{{\cal O}}

\def\CR{{\cal R}}
\def\CS{{\cal S}}
\def\CT{{\cal T}}

\def\bC{{\bf C}}

\def\bP{{\bf P}}

\def\bR{{\bf R}}

\def\bT{{\bf T}}

\def\bZ{{\bf Z}}

\def\p{\partial}
\def\pb{\bar{\partial}}


\def\jb{\bar{j}}

\def\zb{\bar{z}}

\def\Tr{{\rm Tr}}

\def\Det{{\rm Det}}

\noblackbox
\newcount\figno
 \figno=1
 \def\fig#1#2#3{
 \par\begingroup\parindent=0pt\leftskip=1cm\rightskip=1cm\parindent=0pt
 \baselineskip=11pt
 \global\advance\figno by 1
 \midinsert
 \epsfxsize=#3
 \centerline{\epsfbox{#2}}
 \vskip 12pt
 {\bf Fig.\ \the\figno: } #1\par
 \endinsert\endgroup\par
 }
 \def\figlabel#1{\xdef#1{\the\figno}}
 \def\encadremath#1{\vbox{\hrule\hbox{\vrule\kern8pt\vbox{\kern8pt
 \hbox{$\displaystyle #1$}\kern8pt}
 \kern8pt\vrule}\hrule}}

\Title
{\vbox{
 \baselineskip12pt
\hbox{hep-th/0312022}
\hbox{HUTP-03/A078}
\hbox{IHES/P/03/65}
\hbox{ITEP-TH-60/03}
}}
{\vbox{
 \centerline{Quantum Foam and Topological Strings}
 }}
\centerline{Amer Iqbal$^{a}$, Nikita Nekrasov$^{b}$\footnote{$^{\dagger}$}{On leave of absence from: ITEP, Moscow, 117259, Russia},
 Andrei Okounkov$^{c}$
and Cumrun Vafa$^{a}$}
\bigskip
\centerline{$^{a}$ Jefferson Physical Laboratory, Harvard University}
\centerline{Cambridge, MA 02138, USA}
\centerline{}
\centerline{$^{b}$ Institut des Hautes Etudes Scientifiques}
\centerline{Bures-sur-Yvette, F-91440, France}
\centerline{}
\centerline{$^c$ Department of Mathematics, Princeton University}
\centerline{Princeton, NJ 08544, USA}
\centerline{}



\smallskip
\centerline{\bf Abstract}
We find an interpretation of the recent connection found between
topological strings on Calabi-Yau threefolds and crystal melting:
Summing over statistical mechanical configuration of melting crystal is
equivalent to a quantum gravitational path integral involving
fluctuations of K\"ahler geometry and topology.  We
show how the limit shape of the melting crystal emerges
as the average geometry and topology of the quantum foam at the
string scale.
The geometry is classical at large length scales, modified
to a smooth limit shape dictated by mirror geometry at string
scale and is a quantum foam at area scales $\sim  g_s \alpha'$.

\Date{December 2003}

\newsec{Introduction}
The idea that quantum gravity should lead to wild fluctuations
of topology and geometry at the Planck scale is an old idea
\refs{\whee , \hawk}.  Despite the plausible intuitive nature of this
idea it has not become a precise part of our current understanding
of quantum gravity.  Even with the advent of superstring theory
as a prime candidate for a theory of quantum gravity, we still
have a long way to go to understand the geometry of space at short
distances.  For example for type IIA superstrings
in ${\bf R}^{10}$, even for small $g_s$,  we do
not know how the spacetime looks like at the Planck scale $l_p=g_s^{1/4}
l_s$.  Trying
to probe geometries at such a distance scale requires
high center of mass energies which lead to creation
of black holes and the puzzles associated with it.

A good laboratory for studying these questions is the
topological strings.  Topological A-model strings
\wittos\ involves
from the worldsheet viewpoint sums over holomorphic
maps to the target space.   The critical
case corresponds to the target space being Calabi-Yau threefolds. From
the target space
perspective one is considering a gravity theory known
as `K\"ahler gravity' \besa.  Roughly
speaking one is integrating over the space of all K\"ahler metrics
on the Calabi-Yau manifold.  Of course, the action of this gravity
theory is not just the Einstein action, as it would be a total
derivative on the K\"ahler manifold.

Topological string is rich enough
to have many features familiar from superstrings: Lagrangian
D-branes, (known as A-branes), large $N$
open string/closed string dualities \gova\ etc.  Moreover, in the
Berkovits formalism, topological
string amplitudes can
be viewed as part of the worldsheet theory of superstring
(for a recent discussion of this see \bov),
where they are interpreted as  F-term amplitudes
\refs{\bcov , \naret}.
It is thus interesting to ask whether or not we can have a more
clear picture of the gravitational quantum foam in this theory.

Recently a duality was discovered in \orv\ which related
topological string amplitudes in the A-model to a statistical mechanical
partition function involving crystal melting.  Such a duality between a
quantum theory and a classical system has one well known analog: Quantum
field theories can typically be viewed as sums over classical
configurations of fields with some weights determined by the action. It
is thus natural to try to interpret the configurations of melting
crystal as target space field theory of the A-model, which
should be a gravitational theory involving
fluctuations of topology and geometry.  Thus the partition
function of the crystal melting would get mapped to a weighted sum over
various topologies.

\ifig\twophase{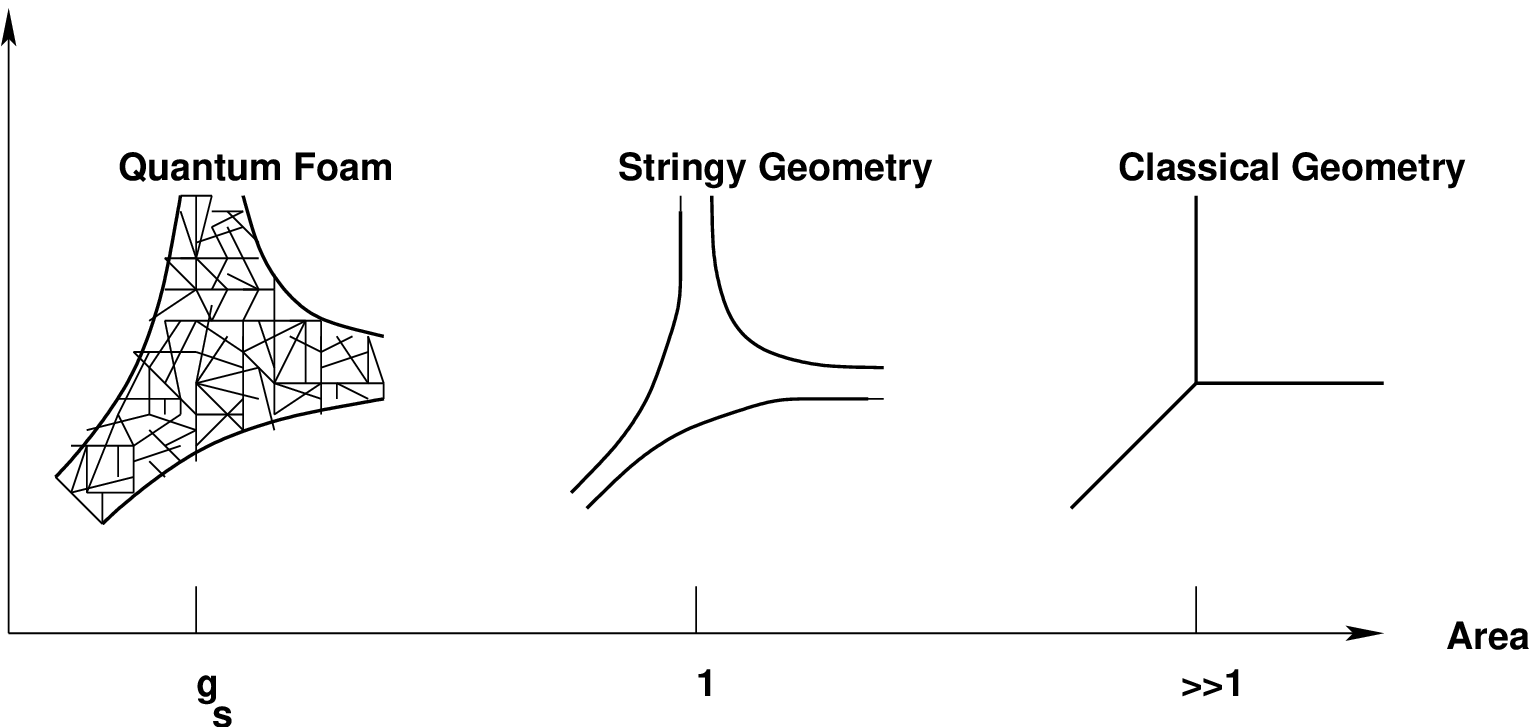}{90}
{The diagram shows how the geometry varies as we change
the area scale, drawn in string units.}

In this paper we find that
this is indeed a consistent interpretation of this duality: We find that
at short distance when the area is of order of $ g_s l_s^2$
where $l_s={\sqrt \alpha'}$ is the string scale, (assuming $g_s<<1$) a
gravitational
quantum
foam takes over.  The foam gets identified with
the fluctuating boundary configuration of the crystal.
The geometry and topology fluctuate
wildly at this scale.  At larger scales, at the string scale,
a smooth geometry takes over which is the average fluctuation
of the foam dictated by geometry of the mirror manifold.  This is
a macroscopic regime for the crystal where a smooth limit shape
emerges for the boundary of the molten crystal.
At yet
larger scales the classical geometry description takes over.
At these scales, the crystal appears not to have molten at all.
See \twophase .  It is also natural to see how this picture
can be embedded in superstrings.  One possible approach
to this idea is currently under investigation \osv.

What we find is that the quantum fluctuations of topology involves blowing
 up the original Calabi-Yau manifold $X$ along
points and holomorphic curves.
 The blown
up space ${\widehat X}$ is no longer Calabi-Yau. It comes canonically with a line
bundle $L$ whose local holomorphic sections are identified with the
local holomorphic functions on $X$, which vanish along the curves and points we have blown up.
The new K\"ahler form on ${\widehat X}$ is obtained from the original one
on $X$ up to the addition
of the first Chern class of $L$ times $g_s$.
Precise definition of summing over ``quantum K\"ahler structures'' consists
of integrating over ``ideal sheaves'', which are ``torsion free sheaves
with vanishing first Chern class''.  In string theory we often deal
with these objects when considering gauge fields supported on D-branes.
More physically, we replace the integration over K\"ahler geometries
and topologies by the path-integral of a
 $U(1)$ gauge theory on $X$ (related to a topological
 twisting of maximally supersymmetric gauge theory).  Thus our proposal is
similar in spirit
to $SL_{2}({\bR})$ and $SL_{2}({\bf C})$ gauge field interpretations of
two and three
dimensional gravities respectively.

The ideal sheaf is the generalization of
the line bundle, and the moduli space of
ideal sheaves is the generalization of the moduli space of instanton solutions in the
6 dimensional gauge theory on $X$.
The torsion free sheaves fail to be a
line bundle in
real codimension four. The Calabi-Yau $X$ can be blown up along this locus
to produce another space ${\widehat X}$. Torsion
free sheaves can be lifted to ${\widehat X}$ where they become line
bundles. There is no one to one correspondence between the sheaves on $X$
 and blown up geometries ${\widehat X}$, and in
some sense there are more sheaves than geometries.
 We interpret this
mismatch as a lesson of quantum string theory for quantum gravity:
{\it The path-integral space for quantum gravity should include classical
topologies and geometries.  However the actual space we integrate over may
well be bigger
than that given strictly by manifolds with arbitrary topology and metric, as happens
for topological strings. Thus we conclude, that for topological
strings a gauge theory
is the fundamental description of gravity
at all scales
including the Planck scale, where it leads to a quantum gravitational
foam}.
A related example of this, as we will note later, is that
of 3d gravity theory and its formulation
in terms of $SL(2,{\bf C})$ Chern-Simons theory \wittencsg .

The plan of this paper is as follows:  In section 2 we present
the basic idea.  In section 3 we discuss aspects of toric
geometry which is relevant for the main class of non-compact
examples studied in this paper. It is also necessary for
understanding the nature of the quantum foam.  In section 4
we apply this to the ${\bf C}^3$ case.  In section 5
we discuss the generalization of this to the case of
toric 3-folds.  Some of the more technical
discussions are presented in section 6.  In particular
in that section we propose more
precisely what K\"ahler gravity is in terms of
 the $U(1)$ gauge theory of a maximally supersymmetric, topologically
twisted theory.  We also
show how the idea of localization for toric geometries works in this
context by
considering certain deformations, similar to the approach taken
for 4d instantons \neksw .   In section 7 we briefly discuss
generalizations to the compact case.

\newsec{The Basic Idea}
Topological A-model involves, from the worldsheet perspective
summing over holomorphic maps to the target \wittos.  The
critical case, which is the case of interest
in this paper, is when the target space is a Calabi-Yau threefold, which
we denote by $X$.  The
(complexified) K\"ahler
class of the target space, represented by a closed $(1,1)$
form $k$, plays a key role:  It weights
worldsheet instantons by $e^{-k_\Sigma}$ where $k_{\Sigma}$
denotes the integral of $k$ over the image of the worldsheet
$\Sigma$.

As far as the target geometry is concerned at the level
of genus 0 worldsheet, i.e., string theory tree level, one
is studying quantum intersection theory.  This means, roughly
speaking, one is asking whether cycles intersect, up to a fuzziness
involving holomorphic spheres.  Thus the target geometry is inherently
`fuzzy' at the string scale.

The target space description of the field theory, should involve
integration over the space of all K\"ahler metrics on the Calabi-Yau,
with a fixed complex structure.  This was studied in
\besa\ following earlier works
\refs{\witcs , \bcov}, where
it was found that one is studying `K\"ahler gravity'.  Part of the
difficulty in implementing \besa\ as a practical
method for computing topological A-model amplitudes is that one expects
that a `stringy' geometry should take over the classical
description of the geometry at distances of order
of string scale, where the worldsheet instantons
will not be suppressed.  Thus the lack of a precise
meaning for what could one mean
by a
`stringy K\"ahler gravity' is an obstacle to this program.
At any rate, it was found that the classical solutions
of the theory is given by K\"ahler forms $k$, and the
action evaluated at these points is given by
$${\CS}={1\over g_s^2}\int_X {1\over 3!} k\wedge k\wedge k\,. $$
On the other hand recent results indicate that the K\"ahler form in
topological strings is quantized.  The first hint
of that emerged in \gova\ where
it was found that topological string on resolved
conifold is equivalent to $U(N)$ Chern-Simons on $S^3$
provided that the K\"ahler class of the blown up ${\bf P^1}$
is identified with $k({\bf P}^1)=Ng_s$.  More recently
\tvtwo\ it was found that quite
generically if one considers $N$ Lagrangian D-branes (known
as A-branes) inside a CY, then the integral of $k$ along 2-cycles
surrounding them jumps by $N g_s$, generalizing the
observation in \gova .  It is thus natural
to suspect that quite generally $k$ should be viewed as quantized
in units of $g_s$ (for quantization of areas in a different context see e.g.,
\loopqg).  In other words we have
$$k=F g_s\,,$$
where $F$, being an
integral $(1,1)$ form, is the curvature of a $U(1)$ bundle over the
Calabi-Yau.

Note that $\int_X k^3/g_s^3$ could be viewed as the index of the $\overline
\partial$ operator coupled
to a gauge field with field strength $k/g_s$\foot{We neglect the gravitational corrections to the index for the moment}.  Thus
we can view ${\CS}$ as counting the net number of holomorphic
sections
of the bundle, up to a factor of $g_s$, i.e.
\eqn\fundr{Z=\sum e^{{\CS}}=\sum {\rm exp}(Ng_s)}
 where
 $N$ is the number (or more precisely the index)
of holomorphic sections
of this bundle and we sum, in a first approximation, over all
line bundles with
connection.
$N$ can also be viewed as the number of states in the Hilbert space, if
we identify
$$\eqalign{{\rm Calabi-Yau}&\longleftrightarrow {\rm phase \ space}\cr
k\ &\longleftrightarrow \ {\rm symplectic\ form}\cr
 g_s\ &\longleftrightarrow \ \hbar\cr} $$
In this context the geometric quantization identifies the states
in the Hilbert space with holomorphic sections of the bundle
whose curvature is $k/g_s$.

It is also interesting to ask whether this Hilbert space has a natural
interpretation directly in topological strings.  Note that viewing Lagrangian
A-branes as magnetic objects, which give rise to $k$ flux, the electric
objects which couple minimally to the connection whose curvature is $k$,
are 1-dimensional Euclidean objects (0-branes in the more conventional
terminology of superstrings).  Namely if the 1-dimensional
branes  bound a disk $D$, their action is
weighted
by
$$exp(-{i\over g_s}\int_D k )\,.$$
That this is well defined independently of the choice of the
bounding disk $D$, follows because $k$ is quantized in units of $g_s$.
These branes can also be viewed as induced by turning
on a $U(1)$ field strength inside a Lagrangian A-brane.  Thus the
partition function of topological A-model seems to be counting
the number of such states for each configuration of K\"ahler moduli.

It is also useful to expand $k$ near a fixed background
geometry $k_0$, as often is needed in topological strings.  In this context
it is natural to write
the fluctuations in terms of the field strength of a $U(1)$
bundle.  It is also natural to fix the K\"ahler class of the macroscopic
geometry;
in other words we write
$$k=k_0+g_s F$$
and require that the
fluctuation has no periods along 2-cycles of the macroscopic K\"ahler
geometry i.e., $\int_{\beta} F=0$
for $\beta\in H_{2}(X,{\bf Z})$.  This  leads to
\eqn\clsactn{{\CS}= {1\over g_s^2} {1\over 3!} \int_X k_0^3 +
{1\over 2}\int_X F \wedge F \wedge k_0 +
g_s \int_X {1\over 3!}F\wedge F\wedge F}
Thus, apart from the constant background piece $k_0^3$, this can be written
as
\eqn\funde{Z=\sum {\rm exp}({\CS})=\sum q^{ch_3} \prod_i Q_i^{\int_{C_i^{\vee}} ch_2}}
where
$$q=e^{g_s} \quad Q_i=e^{-\int_{C_i} k_{0}}$$
where $C_i, C_i^{\vee}$ are the dual bases of $H_{2}(X,{\bZ})$ and $H_{4}(X,{\bZ})$, and $ch_3$ and $ch_2$ denote
the corresponding Chern classes of the $U(1)$ bundle over the Calabi-Yau 3-fold.
Note that the condition on $F$ above means that it
realizes trivial cohomology class. How come the second and the third Chern
characters do not vanish? The resolution of the puzzle is that $F$
corresponds to the singular $U(1)$ gauge field on $X$. To make it
nonsingular one either blows up the space $X$ or relaxes the notion of the
line bundle. In fact, the latter approach is canonically related to the
former, as we shall explain in more detail in the sections $3$ and $7$.
The natural replacement for the holomorphic line bundle ${\CL}$ is the rank one torsion free
sheaf ${\CI}$ with the same first Chern class.\foot{There is exact sequence of
sheaves:
\eqn\exsqnc{0 \longrightarrow {\CI} \longrightarrow {\CL}
\longrightarrow S_{Z} \longrightarrow 0}
where $S_{Z} = i_{*}{\CO}_{Z}$ is the sheaf supported on the submanifold $Z$ of real codimension
four, which is the union of curves and points. From \exsqnc\ one
derives, e.g. for  $Z$, which is a union of curves ${\Sigma}_{\a}$ and points $P_a$:
\eqn\chrncl{{\rm ch}({\CI}) = {\rm ch}({\CL}) - {\rm ch}S_{Z}}
and in particular:
\eqn\chtwo{{\rm ch}_2 ({\CI}) = {\half} c_1 ({\CL})^2 -\sum_{\a} \ P.D.[{\Sigma}_{\a}]}
\eqn\chthree{{\rm ch}_3 ({\CI}) = {1\over 6} c_1({\CL})^3 - \sum_a \ P.D. [P_a] + \sum_{\a} c_{1}({\Sigma}_{\a}) P.D.
[ {\Sigma}_{\a}]}
(we used Calabi-Yau condition).
It is these correction terms, supported on $Z$ which make our formula
\clsactn\ nontrivial.}

{}From quantum gravity point of view the sum in \funde\ should be
over a suitable class of geometries with a suitable measure. What
we will do in this paper is to propose a precise definition of
this sum in terms of a topologically twisted maximally
supersymmetric $U(1)$ gauge theory on the Calabi-Yau.  We find
that for toric cases this localizes to a very simple formula,
leading to the rules for the statistical mechanics of crystal
melting. It is not a priori
 clear why a simple formula such as this should have worked in an
exact way, as we shall find in this paper.

More precisely what we will find is that if we study
toric Calabi-Yau, the above sum could be localized to toric
geometries, with quantized K\"ahler moduli.  From this point
of view we can describe this as a singular K\"ahler moduli
on the original space.  This is somewhat analogous
to instantons on non-commutative ${\bf C}^2$ which can also
be viewed, in some cases, as smooth instantons on certain
blowup of ${\bf C}^2$ \nekb.  In fact
the sum we perform can also be viewed as summing over
3-dimensional analogs of instantons on non-commutative space.
However it is more natural for our case to view this as a K\"ahler
moduli on a smooth blown up geometry.  The main case we will concentrate on
is for that of toric Calabi-Yau 3-folds.
In fact the sum we do is  an {\it enlargement} of
the space of all {\it quantized} K\"ahler moduli on toric geometries. This is also somewhat
analogous to the 3d gravity described in terms of $SL(2,{\bf C})$
Chern-Simons
connection where the relevant space of connections
is an enlargement of the space
of allowed metrics \wittencsg\ (see \guko\ for a recent
discussion of this theory).  For non-toric
geometries we shall also formulate a precise meaning to the above
sum, in terms of mathematical objects known as torsion free sheaves, which are nothing but the ideal sheaves in our case.
This involves the notion of what singular $U(1)$ `instanton configurations'
mean for 3-folds.

We will develop these ideas further, after we discuss some preliminaries
of toric geometries in the next section.  In section 4 and 5
we assume a simple localization holds and give an interpretation
of the result of \orv .    Physically the localization
should be viewed, as we will note below, as the geometry seen by brane
probes which
are themselves toric.
In section 6 we define more precisely
what the gauge theory  is and why it localizes.

\newsec{Toric Geometry Preliminaries}

It is crucial for this paper to have a simple intuitive grasp
of toric geometries.  To this end here we present a self-contained
introduction
to certain aspects of toric geometries relevant for this paper.

Let us start with the 1-dimensional complex plane ${\bf C}$.  Let us denote
its coordinate by $z$.  There is a $U(1)$ action which corresponds
to phase multiplication.  We can view ${\bf C}$ as
$$z=|z| e^{i \theta}\,.$$
The standard K\"ahler form on ${\bf C}$ can be viewed
as
$$k=i dz\wedge d{\overline z}=d\theta \wedge d|z|^2 $$
Let us define
$$p=|z|^2\,,$$the moment map, in terms of which the K\"ahler form is given by
$$k=d\theta \wedge dp$$
We can view ${\bf C}$ as a circle, parameterized by $\theta$,
fibered over positive half-line $p\geq 0$.

Let us view ${\bf C}$ as the phase space
with $k$ as the symplectic
form.  Furthermore suppose we were to write a basis for the Hilbert
space for the quantization of this phase space.   Motivated
by our discussion of the previous section, we introduce
the Planck constant $\hbar =g_s$ and view $\theta$ as the space coordinate,
$$[\theta , p]=i g_s\,.$$
A basis for the Hilbert space can be taken to be
$$\psi_n(\theta) ={\rm exp}(i n \theta)\,,$$
which leads to
$$p=n g_s.$$

\ifig\har{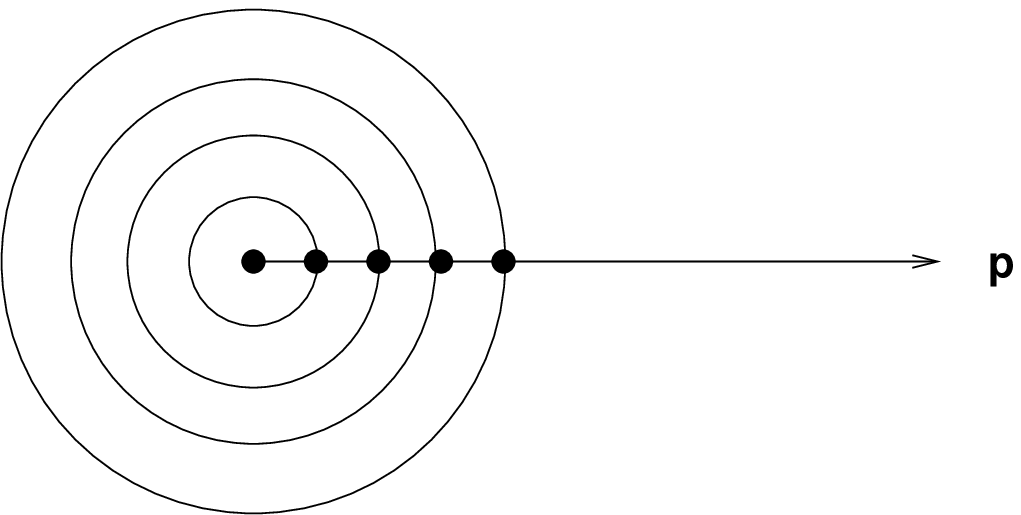}{90}
{The diagram shows how the $p=|z|^2$ half-line can be chosen
as the Lagrangian subspace of complex plane. The states of the
Hilbert space are identified with $p=n g_s$ for positive integer $n$.}

Here $n$ should be integer, because of the periodicity of $\theta$
and $n\geq 0$, by the fact that the momentum $p\geq 0$.  We can identify
the space of states as integer lattice points in the positive
half-line parameterized
by $p$, separated by lattice spacing $g_s$.  The above result
of an integer spectrum for $p=|z|^2$ is familiar in the context
of one dimensional Harmonic oscillator, with the identification of
$H=p=|z|^2$.

We can also
get the same set of states by geometric quantization.  What this
means is that we view $k$ as the curvature of a $U(1)$ bundle, up
to a factor of $\hbar=g_s$
$$k=g_sF,$$
and we are looking for holomorphic sections of the bundle given
by
$${\overline D_A}\psi=({\overline \partial}+{\overline A})\psi =0$$
In this case ${\overline A}=z$ and so a basis for the solutions
are given by
$$\psi_n= z^n e^{-z{\overline z}}.$$
Note that the wave function $e^{in \theta}$ corresponds to the
phase in the $z^n$ factor above.
Both of these viewpoints will be useful for us.

Note that holomorphic functions on ${\bf C}$ act on the Hilbert
space.  This is generated by functions of $z$ and we have
$$z\psi_n =\psi_{n+1}$$
This is a general fact about geometric quantization:
$$[z,{\overline D_A}]=0$$
and thus holomorphic functions are realized as natural
operators on the Hilbert space.

The generalization of this to $n$-dimensional complex plane
is straight-forward.  The space of states in this case
can be identified with an $n$-dimensional crystal sitting
at positive integral points of $(p_1,...,p_n)$ separated
by $g_s$.  In the case of $n=3$ this gives a three-dimensional cubic
crystal, filling the positive octant (which can also be identified
with the states of a three dimensional harmonic oscillator).

So far we have considered the simplest toric geometries.  For
more interesting ones we add more coordinates and impose identifications
in the holomorphic terminology, or constraints modulo gauge equivalences
in the linear sigma model terminology \lins.
In general we will be considering $n+r$ variables $z_i$, where
$i=1,...,n+r$, with
$r$ identifications/constraints.  The identifications/constraints
are encoded in terms of $r$ sets of integral charge vectors $Q^a_i$ where
$a=1,...,r$ and $i=1,...,n+r$.  In the holomorphic description
we identify
$$z_i \sim \lambda^{Q^a_i}_a \ z_i$$
for each $a$, where $\lambda_{a} \in {\bf C}^{*}$ and certain loci are deleted.
In the linear sigma model set up we consider the $r$ constraints
\eqn\cons{\sum_i Q^a_i |z_i|^2=\sum_i Q^a_i p_i=t^a}
modulo $U(1)^r$ gauge symmetry given by
$$z_i\rightarrow e^{i\theta^a Q^a_i}\
z_i.$$
The above constraints \cons\ are linear in the $p_i$ variables.
The space ${\Delta}(X)$ of $p_i$ consistent
with the above equations is generically  $n$ dimensional
subspace of ${\bf R}^n_{\geq 0}$.  It is also easy to see that this
subspace is convex:
If $p^1_i>0$ and $p^2_i>0$ satisfy the charge constraints,
then $\lambda p^1_i+(1-\lambda)p^2_i$ are positive and also
satisfy the constraints for $1\geq \lambda \geq 0$. The space ${\Delta}(X)$
is the quotient $X$ by the torus ${\bT}^n$, which is in turn the quotient
${\bT}^{r+n}/{\bT}^r$.

In the above $t^a$ denote the various K\"ahler moduli of the toric
geometry.  We would also like to consider these spaces as phase
spaces and do quantum mechanics on them.  In doing so we find
that for there to be a well defined wave functions we need
the K\"ahler moduli $t^a$ to
be quantized in units of $g_s$.  This is because if we consider
the wave functions as a function of periodic spatial
coordinates $\theta_i$, single valuedness requires
$$\psi (\theta_i)={\rm exp}(i N_i \theta_i)$$
where $N_i$ are integers.  However $N_i$ cannot
be arbitrary, because the corresponding momenta
$p_i=N_i g_s$ should satisfy \cons :
\eqn\ancon{\sum_i Q^a_i N_i g_s =t^a}
For this to have solutions, we need $t^a$ to
be quantized in units of $g_s$:
$$t^a=m^a g_s$$
which leads to
\eqn\fincon{\sum_i Q^a_i N_i=m^a}
The set of points $\{ N_i\} \subset {\bZ}_{\geq 0}^{n+r}$ satisfying
\fincon\ label
the vector space which span the Hilbert space ${\CH}$ corresponding
to quantization of the space.  They could also be viewed
as the space of sections of the holomorphic bundle with
curvature $k=F g_s$, in case it is sufficiently positive.

\subsec{Examples}
In order to illustrate these ideas we will construct
a few examples.

\ndt${\bf P}^{1}$:  First consider a compact example of dimension
one, namely ${\bf P}^1$.  This can be realized by two variables
$(z_1,z_2)$ with one charge vector $(1,1)$.  Let $t$ denote
the K\"ahler class.  Then we have
$p_1+p_2=t$

\ifig\pone{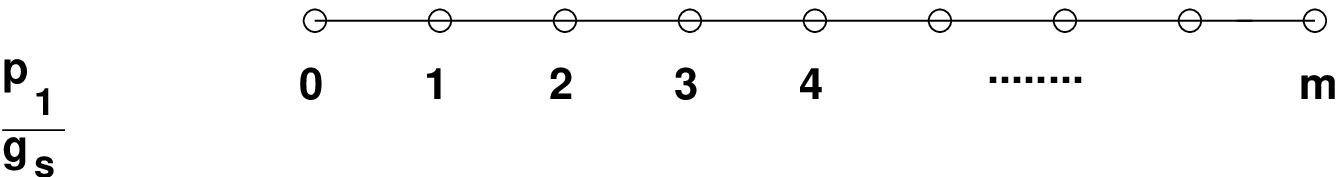}{90}
{The toric diagram for ${\bf P}^1$ is an interval.  The
sphere arises by recalling that over each point there is a circle
which shrinks at the two ends.
The diagram shows how the states in the quantum mechanical
problem corresponds to integer point in $g_s$ units.}

If we take $t$ to be quantized in $g_s$ units, $t=mg_s$ then we can
identify the states of the Hilbert space with integer
points $N_1\geq 0, N_2\geq 0$, subject to
\eqn\pa{N_1+N_2=m\ .}
Note that $N_2=m-N_1>0$ implies that we can identify this space with
lattice points on the positive line satisfying
$$0\leq N_1\leq m$$
In particular this space has $m+1$ points, see \pone .
The corresponding
K\"ahler class $k$ is identifiable with $g_s$ times the
curvature corresponding to the line bundle $O(m)$ over
${\bf P}^1$.  This is the bundle whose holomorphic
sections can be identified with homogenous degree $m$ polynomials
in $z_1, z_2$. A solution of \pa , $(N_{1},N_{2})$, corresponds to a
monomial $z_{1}^{N_{1}}z_{2}^{N_{2}}$.  Thus degree $m$ monomials are
 naturally identifiable with the above
$m+1$ points.

\ndt${\bf C}^{2}$:
As our next example we consider a non-compact 2-dimensional
space.  This will correspond to blowing up ${\bf C}^2$ at a point.
This means we take a point on ${\bf C}^2$ and replace it with a
${\bf P}^1$.  The geometry near this ${\bf P}^1$ is ${\cal O}(-1)\rightarrow
{\bf P}^1$, and in particular it is not a Calabi-Yau space.  We take
three variables and take one charge
$$Q=(1,1,-1)\,.$$
Then the momenta satisfy
$$p_1+p_2-p_3=t\,.$$
The ${\bf P}^1$ is identified with the subspace above with
$p_3=0$, and the normal bundle ${\cal O}(-1)$ over ${\bf P}^1$
is identified with varying $p_3\geq 0$.
Again if we quantize $t$ we get for the allowed integral
momenta the condition that $N_i\geq 0$ and
$$N_1+N_2-N_3=m\,.$$
Note that this can be identified with the space of positive
integral points in the quadrant $(N_1,N_2)$ with the further
restriction that
$$N_3=N_1+N_2-m\geq 0\,.$$

\ifig\poneb{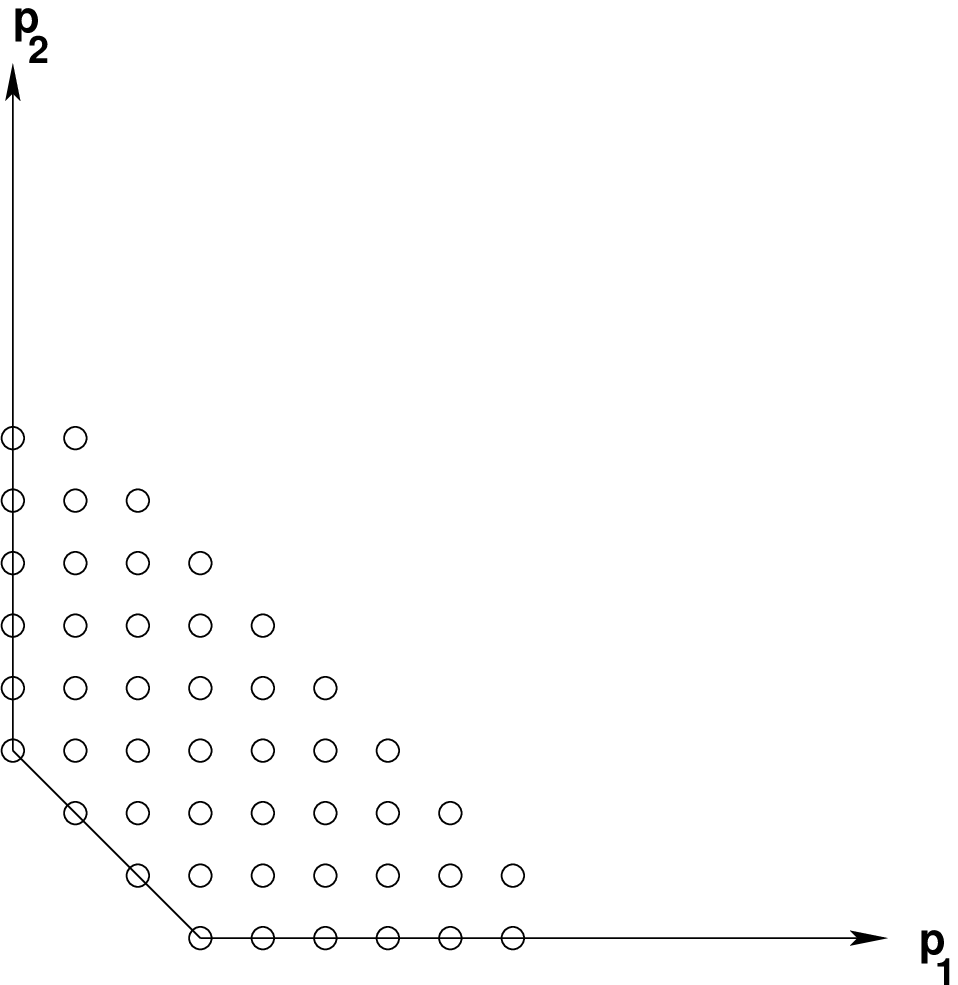}{90}
{The diagram shows blowup of a point of ${\bf C}^2$ and replacing
it with a ${\bf P}^1$.  The case depicted in the figure has
K\"ahler class $t=3g_s$.}

The K\"ahler class over ${\bf P}^1$ corresponds
to the ${\cal O}(m)$ bundle, as discussed before.  This
is called blowing up a point in ${\bf C}^2$ because
we have removed the origin and replaced it by a ${\bf P}^1$,
which is represented by an interval in the plane.
 Note
that compared to ${\bf C}^2$ we have deleted
a triangle of points from the space, by blowing up the
point at the origin.  In other words we have
$m(m+1)/2$ less elements in the Hilbert space
(with a natural definition of `counting' the states
of the Hilbert space).

One can also blow up to a singular space.  For example
let us consider a singular blow up of ${\bf C^2}$:  consider
three variables
with one charge
$$Q=(2,1,-1)\,.$$
Then we have
$$2p_1+p_2-p_3=t\,.$$
The corresponding space has a ${\bf Z_2}$ singularity
at $p_1=t/2 ,p_2=0, p_3=0$.  To see this
note that the ${\bf C}^*$ action generated
by $Q$ is gauge fixed by going
to fixed value of $z_1$.  But
this leaves an extra $Z_2$ part of the
action  which sends $(z_2,z_3)\rightarrow (-z_2,-z_3)$.
Note again in this example if we wish an integral subspace
of states we can take $t=m g_s$ leading to the positive integral
momenta constrained by
$$2N_1+N_2-N_3=m\,.$$

We can also blow up more than one point on ${\bf C}^2$.  For example
consider four variable with two charges
$$Q^1=(1,1,-1,0)\,,$$
$$Q^2=(1,0,1,-1)\,.$$
Giving
$$p_1+p_2-p_3=t^1\,,$$
$$p_1+p_3-p_4=t^2\,.$$
This space corresponds to the subspace of the positive $(p_1,p_2)$
quadrant satisfying
$$p_1+p_2\geq t^1\,,$$
$$2p_1+p_2\geq t^1+t^2\,.$$

\ifig\twob{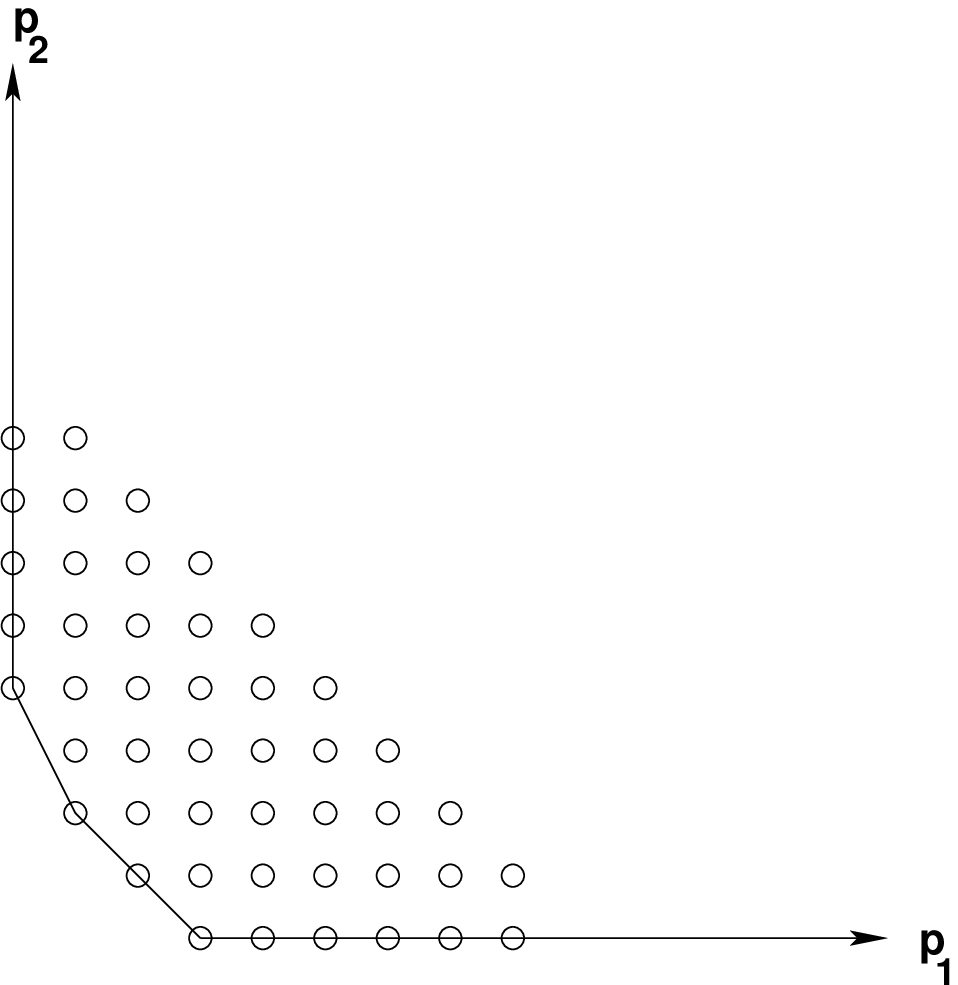}{90}
{The diagram shows two blow ups of ${\bf C}^2$.
The diagram shown corresponds to $t^1=3 g_s$ and $t^2=g_s$.}

Again, if we wish to obtain integral points we can consider
quantizing $t^1,t^2$ in units of $g_s$ leading to
$$N_1+N_2-N_3=m^1\,,$$
$$2N_1+N_2-N_4=m^1+m^2\,.$$
See \twob .
Clearly we can consider more blow ups.  In this way
we can obtain arbitrary convex subset of integral points
of ${\bf C}^2$ kept.

\ndt${\bf C}^{3}$: For our next example consider ${\bC}^3$ blown up at a point.  This
corresponds to removing a point in ${\bf C}^3$ and replacing it with a ${\bf P}^2$.  The embedding of ${\bf
P}^2$ in the space gives ${\cal O}(-1)\rightarrow {\bf P}^2$. Note that again
this is not a Calabi-Yau geometry (as the normal bundle is not ${\cal O}(-3)$).
This can be realized by considering 4 variables with one charge
 $$Q=(1,1,1,-1)\,,$$
 leading to
$$p_1+p_2+p_3-p_4=t\,,$$
 which can be identified with the positive octant $(p_
 1,p_2,p_3)$ with
the corner removed:
$$p_4=p_1+p_2+p_3-t\geq 0.$$

\ifig\ppp{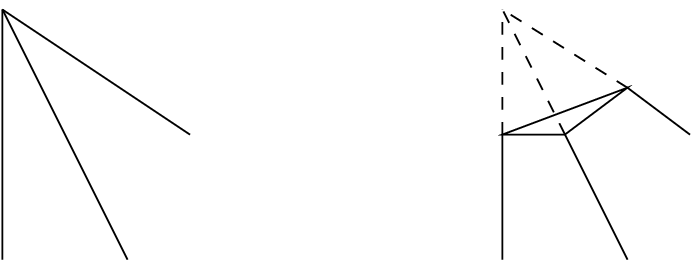}{90}
{Blowing up the origin of ${\bf C}^{3}$ corresponds to chopping off the
corner of the base of ${\bf C}^{3}$ and replacing it with a triangle.}

 The corner is replaced by the triangle
 $$p_1+p_2+p_3=t$$

which corresponds to ${\bf P}^2$ of size $t$ as shown in \ppp.

\ifig\twophase{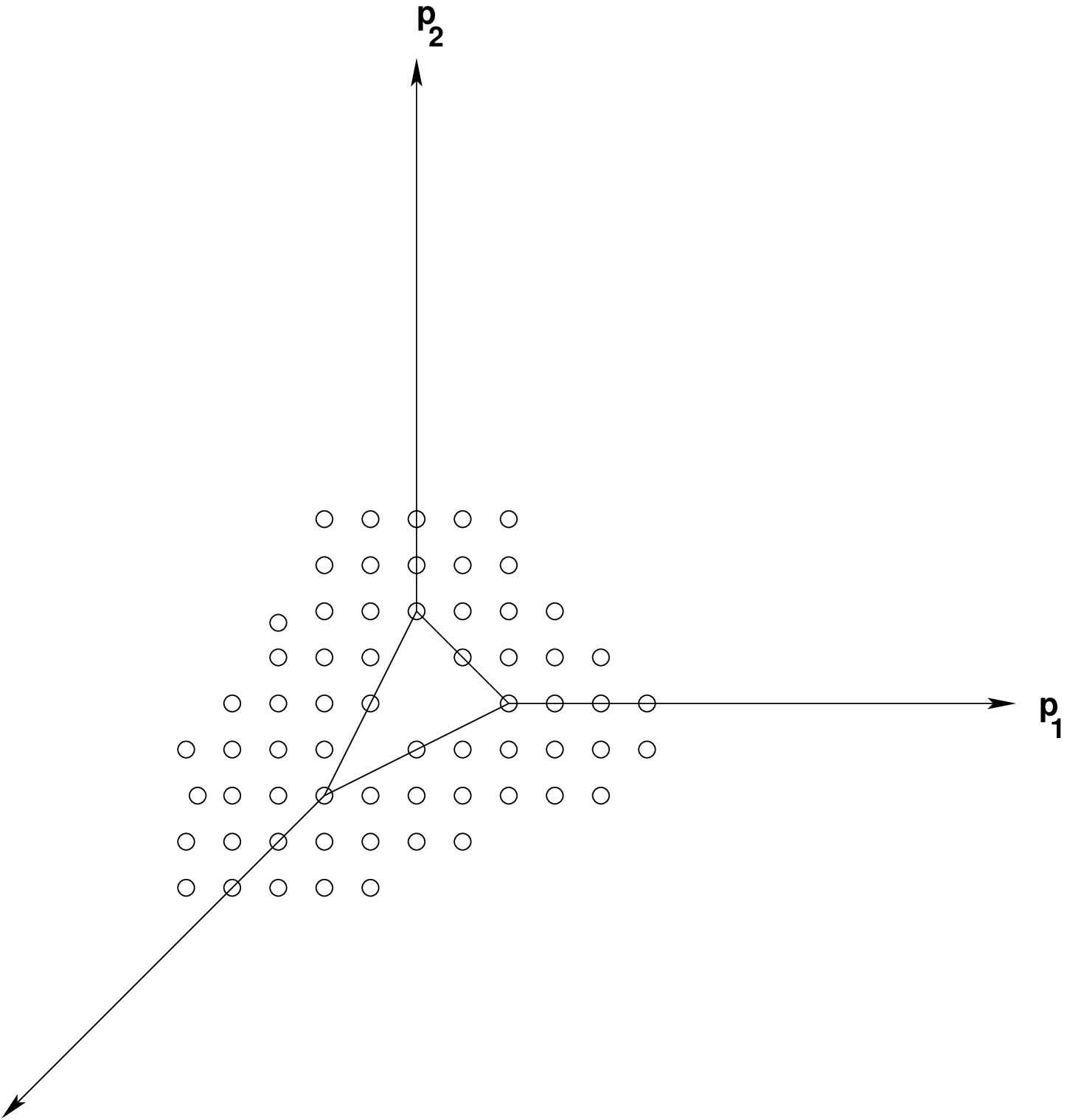}{90}
{The diagram shows the blowing up of a point in ${\bf C}^3$
and its replacement with ${\bf P}^2$.  The case depicted
corresponds to the K\"ahler class $t=2g_s$.}

Again if we consider $t$ to be quantized we have for the allowed quantum
states with $N_i \geq 0$:
$$N_1+N_2+N_3-N_4=m$$
Compared to ${\bf C}^3$ we have deleted $m(m+1)(m+2)/6$ points (states)
from the Hilbert space (see \twophase ). We shall meet this example below, in the context of
noncommutative instantons.

For our next example consider two blow ups of ${\bf C^3}$:  We first
blow a point up to ${\bf P}^2$ and then blow up a point on the
blown up ${\bf P}^2$ to a ${\bf P}^1$.  This can be described by 5 variables with two
set of charges
$$Q^1=(1,1,1,-1,0)\,,$$
$$Q^2=(1,0,0,1,-1)\,,$$
which leads to
$$p_1+p_2+p_3-p_4=t^1\,,$$
$$p_1+p_4-p_5=t^2\,.$$
\ifig\pf{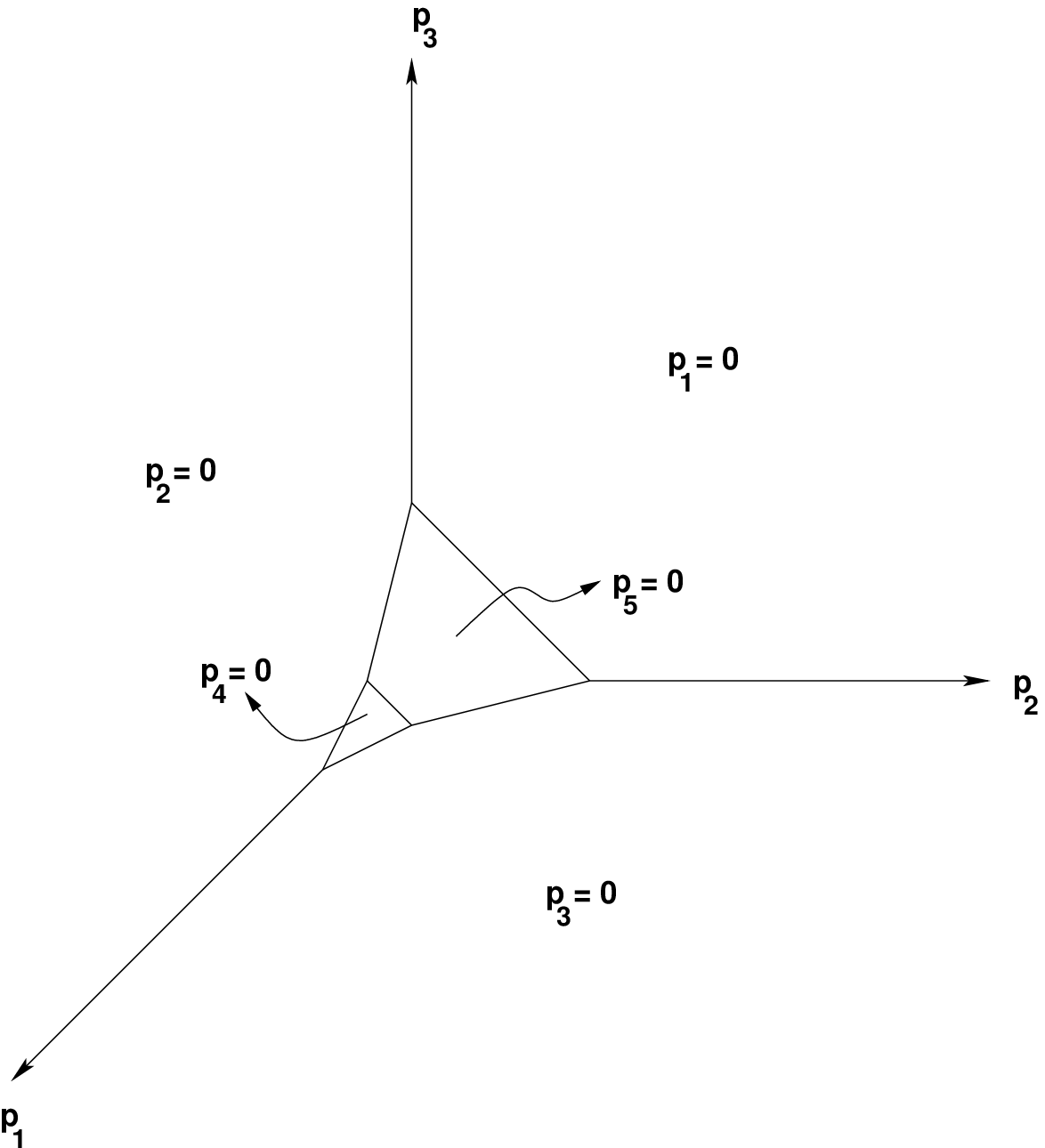}{90}
{The diagram shows two blow ups of ${\bf C}^3$
leading to a ${\bf P}^2$ and Hirzebruch surface ${\bf F}_1$.}

This geometry is depicted in \pf .
If $t_1>t_2$ this corresponds to the geometry of a ${\bf P^2}$
connected to a Hirzebruch surface ${\bf F_1}$,
inside the blown up ${\bf C}^3$.
Again if we consider the integral subspace
the states we keep can be easily deduced.

\ifig\multi{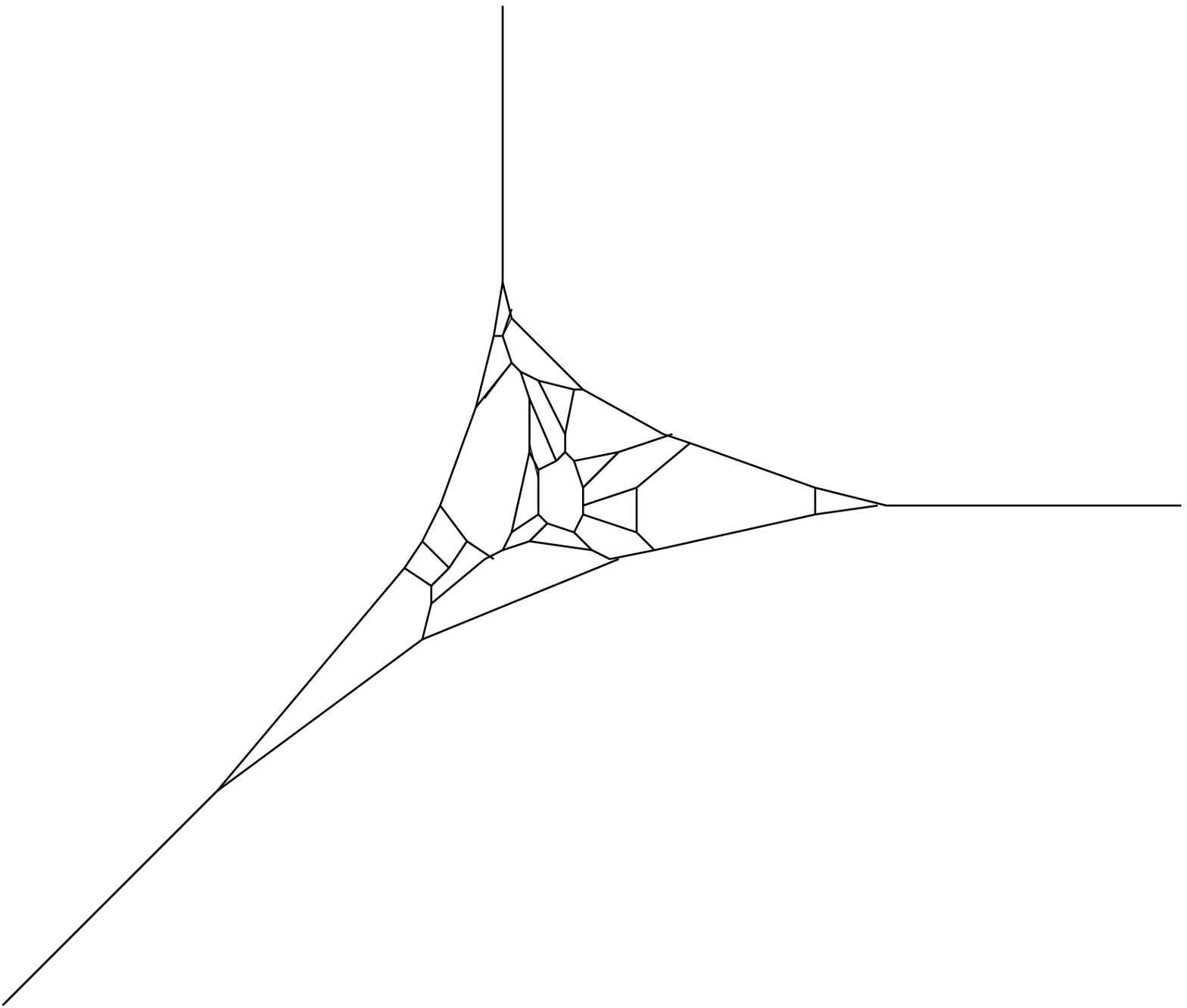}{90}
{We can also consider more general toric blowups of ${\bf C}^3$.}

Clearly one can continue this to more complicated blow ups.
One such example is shown in \multi.

${\cal O}(-1)\oplus {\cal O}(-1) \rightarrow {\bf P}^{1}$:
 We can also consider 3d toric geometries which are not obtained by
blowing up ${\bf C}^3$. For example we can consider four variables with
one charge
$$Q=(-1,1,-1,1)\,,$$
leading to
$$-p_1+p_2-p_3+p_4=t\,.$$
This can be viewed as the subspace of the octant $(p_1,p_2,p_3)$ with
the further constraint
$$p_4=p_1-p_2+p_3+t \geq 0\,.$$

\ifig\res{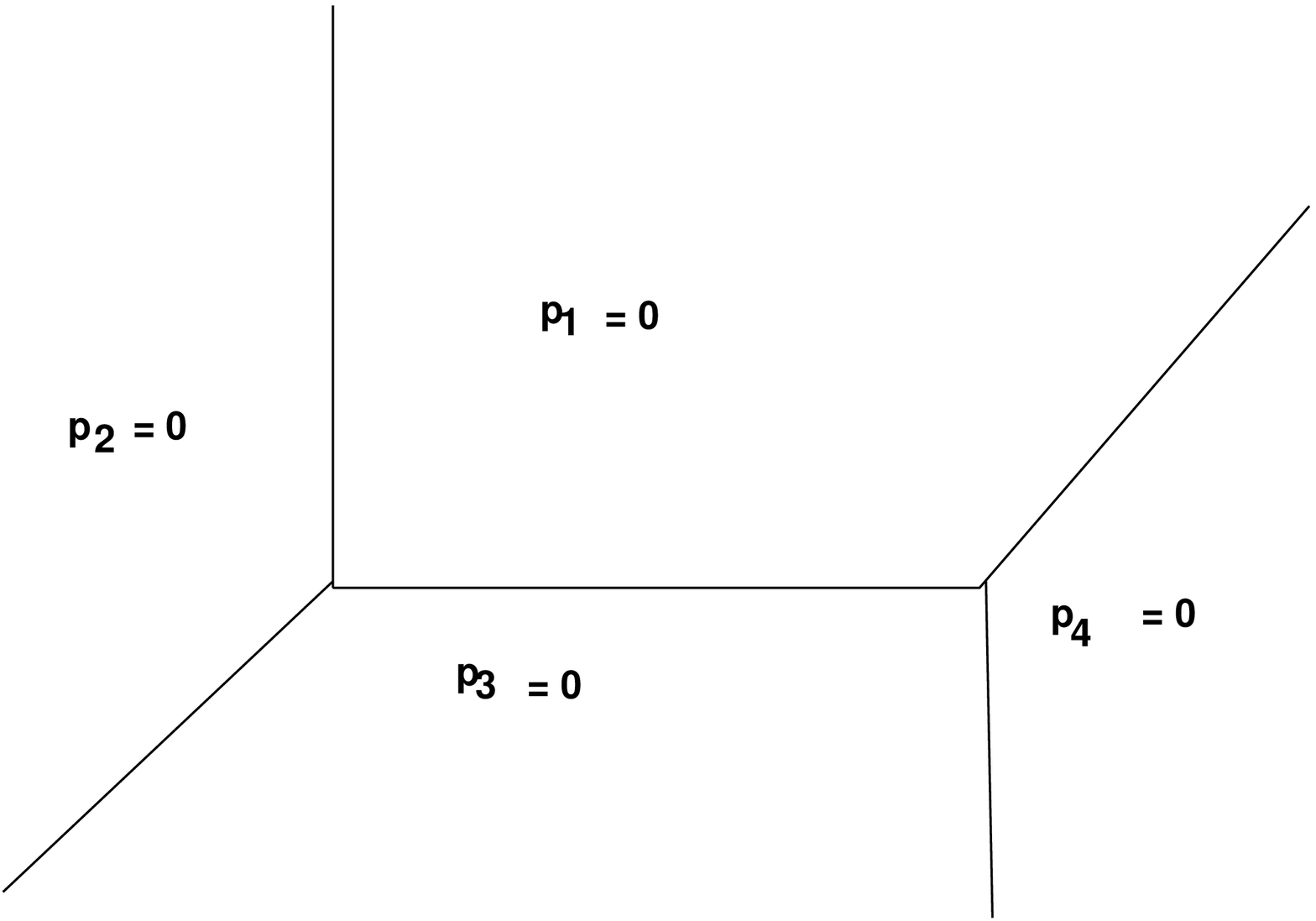}{90}
{The diagram shows the toric geometry of
${\cal O}(-1)\oplus {\cal O}(-1)\rightarrow {\bf P}^1$.}

This geometry is depicted in \res .
Note that
$p_1=p_3=0$ in this space leads to the projection of ${\bf P}^1$ of size
$t$, parameterized by $(p_2,p_4)$ in the momentum space. This in fact
corresponds to $O(-1)\oplus O(-1)\rightarrow {\bf P}^1$ geometry, which
is a Calabi-Yau space. We can take $t=mg_s$ and obtain points of the
Hilbert space with $N_i\geq 0$ satisfying
$$N_1+N_2-N_3-N_4=m\,.$$

 We can also blow up
points in this space.  To do so we introduce one more variable (neutral
under the previous charge) and consider the additional charge
$$(1,0,1,1,-1)$$
which blows up the point $p_1=p_3=p_4=0$.  Or we
could also blow up along ${\bf P}^1$.  For example we consider instead
the additional charge
$$(0,0,1,1,-1)$$

\ifig\ponem{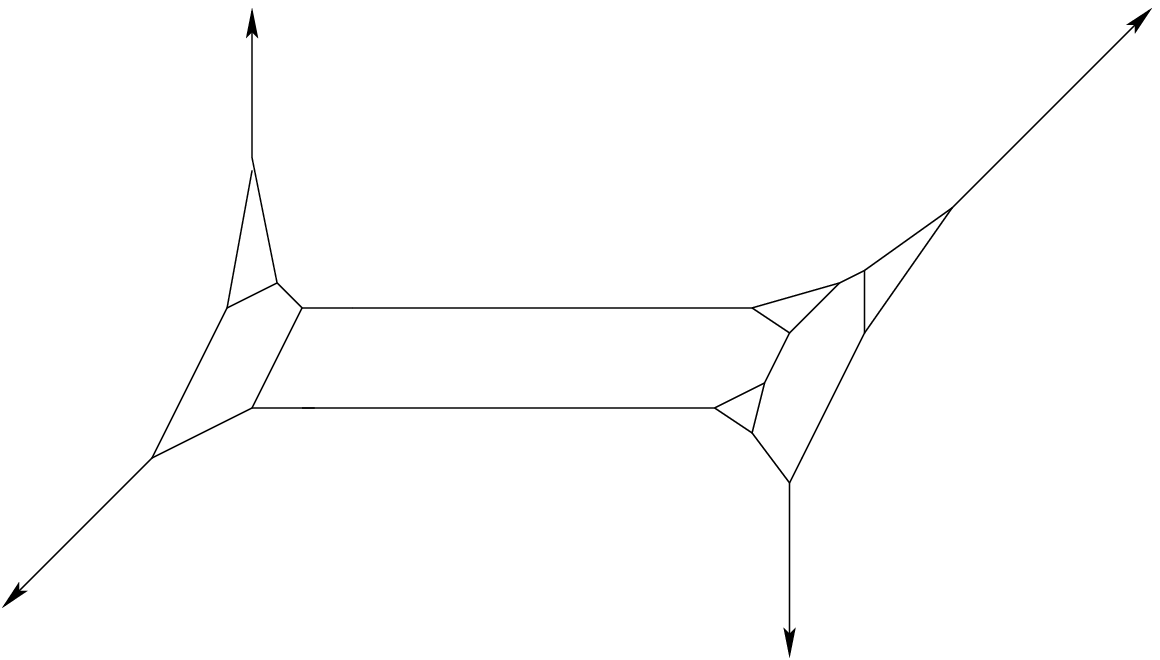}{90}
{The diagram shows blow ups of $O(-1)\oplus O(-1)\rightarrow {\bf P}^1$
along points at the vertices and ${\bf P}^1$ (along the edge).}

We can also consider both 2 dimensional blow ups, i.e.,
blow ups along ${\bf P}^1$ and 3 dimensional
blow up of a point.  We can also consider
various multi-blow ups. See as an example \ponem .

\subsec{Toric Branes} We will also be interested in Lagrangian A-branes which can be
described by toric geometry.  These were first studied in this context in
\mic\ : For simplicity we only discuss the case of ${\bf C}^3$.  Though
this can be extended to other toric geometries, as discussed in \mic : Defining the
image of a Lagrangian A-brane in the base of the toric fibration, i.e.,
$(p_1,p_2,p_3)$ space, uniquely fixes its dependence on the angular part.  An
interesting class of A-branes was studied in \mic\ which has the geometry given by
$$p_1=p_2+a=p_3+a\,, \quad \theta_1+\theta_2+\theta_3=0$$
\ifig\abrane{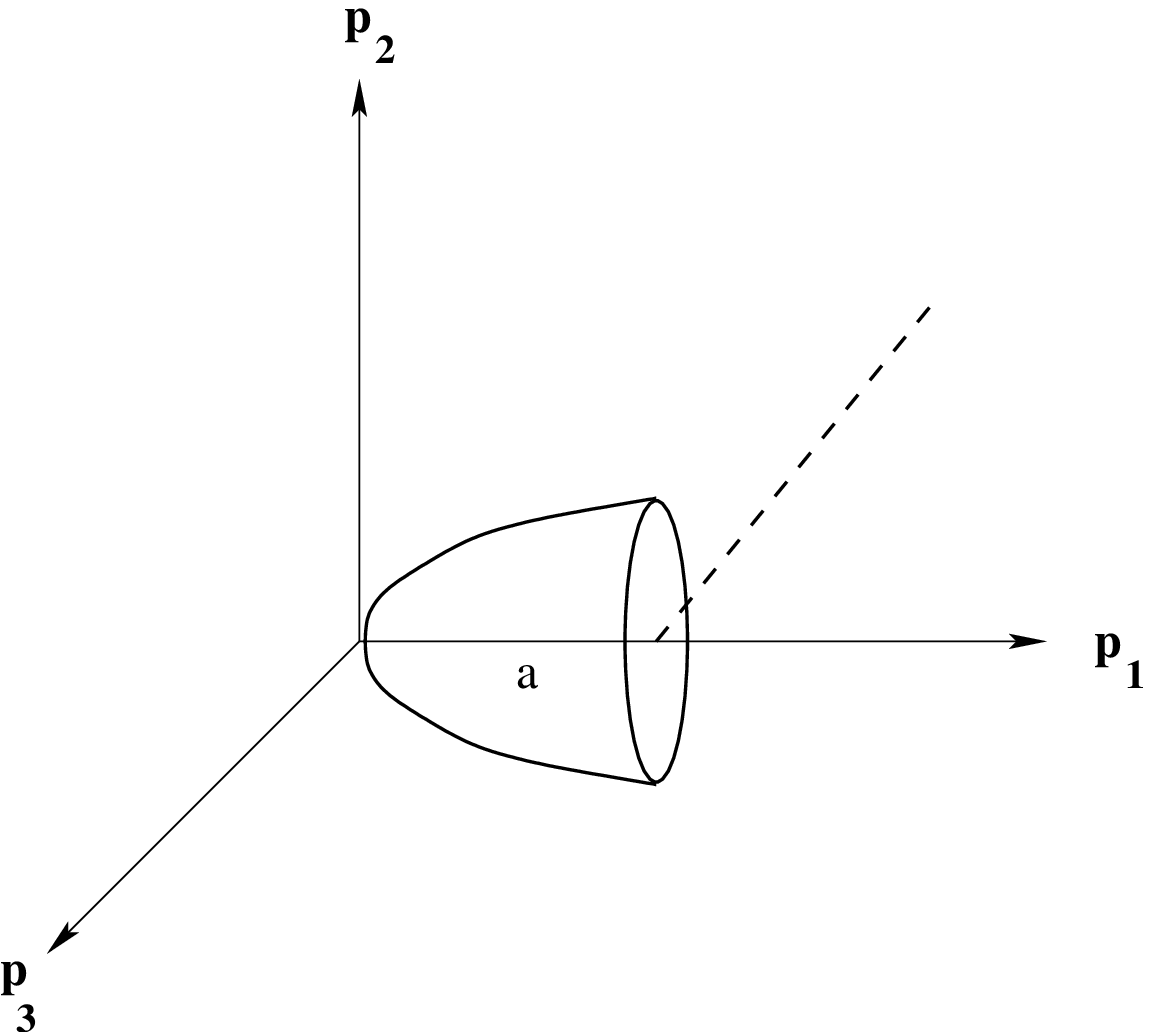}{90}
{The diagram shows A-brane (represented by dashed line)
which projects to a half-line in the toric base.  Its topology
is ${\bf S}^1\times {\bf R}^2$ and there is a minimal disc with area $a$
ending on the brane (the circle comes from the fiber).}

where $a>0$ (note that $a$ has dimensions of area, as $p_i=|z_i|^2$).
These branes have the topology of
${\bf S}^1\times {\bf R}^2$ and end on the $p_1$ axis.
The moduli of these branes is fixed by the parameter $a$ which combines
with the expectation value of the Wilson loop on ${\bf S}^1$ to become
a complexified parameter.  See \abrane .
There is a holomorphic disc in the geometry
which ends on the Lagrangian submanifold. This disc projects
onto the interval $0\leq p_1\leq a$, $p_2=p_3=0$.  The
area of this disc is $a$.  We can view this brane as
the probe of ${\bf C}^3$ geometry.

\subsec{Twisted masses}

In the linear sigma model construction of the toric Calabi-Yau $X$ one can
introduce the so-called twisted masses \refs{\hh , \hv}, which
deform the sigma model by some potential terms
related to holomorphic isometries. In principle, one can turn on the
mass for each chiral field $z^i$, $i=1,\ldots, n+r$. We shall denote them by
${\e}_i$.
In what follows we shall use the following vector field on ${\bC}^{n+r}$ which descends to $X$:
\eqn\omg{{\Omega} = \sum_{i=1}^{n+r}\epsilon_{i}{\partial \over \partial \theta_{i}}}
$$={i\over 2} \sum_{i=1}^{n+r} {\e}_i \left( z^i {\p}_i - {\zb}^i
{\pb}_i \right)$$
Where $\theta_{i}$ is the phase of $z_{i}$. The vector ${\partial \over \partial \theta_{i}}$
 rotates the
phase of $z_{i}$. Hence the fixed locus of this vector is where $z_{i}=0$.
For generic masses the generic orbits of this vector field densely fill
Lagrangian three-tori, which are the fibers of the projection:
\eqn\mmnt{{\vec p} : X \to {\Delta}(X)}
The fixed points $f$ of this vector field
are the vertices of ${\Delta}(X)$. Near each such vertex $X$ looks like a
copy of ${\bC}^n$, and ${\Omega}$ looks like:
\eqn\lclomg{{\Omega} \to {\Omega}_f = \sum_{\alpha=1}^{n}\epsilon_{\alpha f}{\partial\over \partial
\theta_{\alpha}}}
$$={i\over 2} \sum_{{\a} = 1}^{n} {\e}_{{\a}f} \left( z^{\a} {\p}_{\a} -
{\zb}^{\a}
{\pb}_{\a} \right)$$
where ${\e}_{{\a}f}$ -- the local weights -- are some
linear functions of ${\e}_i$ and $z_{\alpha}$ are the local coordinates in terms of which
the vertex is given by $z_{\alpha}=0$. $\theta_{\alpha}$ are the phases of $z_{\alpha}$.
For future use let us recall that the vector field $\Omega$ on the K\"ahler
manifold $X$, with K\"ahler form $k_{0}$, is symplectic and is generated by some Hamiltonian $H$:
\eqn\ham{\iota_{\Omega} k_0 = dH, \qquad H = \sum_i {\e}_i p_i}

Two fixed points $f_1$ and $f_2$ may be connected by an ${\Omega}$-invariant
two-sphere ${\bP}^1$, which we sometimes denote by ${\bP}^1_{e}$, where $e =
(f_{1}, f_{2})$. Changing the order $f_1 \leftrightarrow f_2$ does not
change the sphere geometrically, but reverses the orientation of the
rotation the vector field $\Omega$ induces on it. Thus one ends up with the
graph ${\Gamma}$ with vertices $f$ and edges $e$, each edge being endowed
with the pair of integers $m_1$, $m_2$, describing the topology of the
normal bundle to the sphere. Of course, the graph $\Gamma$ is formed by the
edges and the vertices of the (noncompact) polytope  ${\Delta}(X)$ which is
the image of $X$ under the ${\bT}^n$-moment map.

Here are a few  examples:

\ndt{\bf ${\bf C}^{2}$:} Let us denote by $(z_{1},z_{2})=(|z_{1}|e^{i\theta_{1}},|z_{2}|e^{i\theta_{2}})$
the coordinates of ${\bf C}^{2}$. In this case the vector field is given by
$$\Omega=\epsilon_{1}{\partial \over \partial \theta_{1}}+\epsilon_{2}{\partial \over \partial \theta_{2}}.$$
Its action at a point $(z_{1},z_{2})$ is given by
$$(z_{1},z_{2})\rightarrow (z_{1}e^{i\epsilon_{1}},z_{2}e^{i\epsilon_{2}}).$$
Thus ${\bf C}^{2}$ has only one fixed point given by $(z_{1},z_{2})=(0,0)$. There are also
two fixed planes under $\Omega$ given by $(z_{1},z_{2})=(0,z_{2})$ and $(z_{1},z_{2})=(z_{1},0)$.
\ifig\ctwofixed{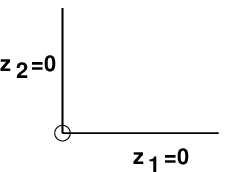}{30}
{The fixed point locus of ${\bf C}^{2}$. The small circle shows the fixed point in the geometry.}
 Thus
on the $(|z_{1}|^{2},|z_{2}|^{2})$ plane we can represent ${\bf C}^{2}$ as shown in \ctwofixed.

\ndt{\bf ${\bf P}^{1}$.} The linear sigma model charges for
this geometry are $(1,1)$. This
means that we can represent ${\bf P}^{1}$ as the quotient of ${\bf C}^{2}$ by the
following action:
$$(z_{1},z_{2})\rightarrow (st_1 z_{1},st_2 z_{2})\ ,\qquad s \in {\bC}^{*}.$$
The vector field $\Omega=\epsilon_{1}{\partial \over
\partial\theta_{1}}+\epsilon_{2}{\partial \over \partial \theta_{2}}$
descends to a vector field $\widehat{\Omega}$ on ${\bf P}^{1}$. In the coordinate patch $z_{1}\neq 0$ ($z_{2}\neq 0$)
the local coordinate on ${\bf P}^{1}$ is ${z_{2}\over z_{1}}=z=|z|e^{i\theta}$ ($z_{1}/z_{2}=w=|w|e^{i\phi}$) and
the vector field $\widehat{\Omega}$ is given by
$$\widehat{\Omega}={\partial \over \partial \theta}\ ,\qquad z_{1}\neq 0,$$
$$\widehat{\Omega}={\partial \over \partial \phi}\ ,\qquad z_{2}\neq 0.$$
\ifig\ppone{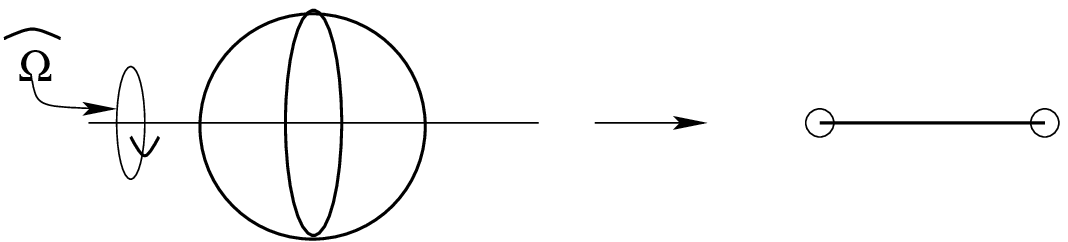}{90}
{The vector field $\widehat{\Omega}$ rotates the ${\bf P}^{1}$ around the axis passing through $z_{1}=0$ and $z_{2}=0$.}
It is clear that the two fixed points of the vector field $\widehat{\Omega}$
are given by $z_{1}=0$ and $z_{2}=0$, \ppone. The ``weights'' of the toric action are given by
the action of $\widehat{\Omega}$ on the tangent space to the fixed points. In this case the weights are given
by $(\epsilon_{1}-\epsilon_{2})$ and $(\epsilon_{2}-\epsilon_{1})$ corresponding to the fixed points
$z_{2}=0$ and $z_{1}=0$ respectively. This follows from the fact that at the fixed point given by $z_{1}=0$
the vectors in the tangent space under the toric action given by
$$z_{2}\rightarrow s t_{2}\ z_2$$
transform as $v\rightarrow ve^{i(\epsilon_{2}-\epsilon_{1})}$.

\ndt{\bf Local ${\bP}^1$ in Calabi-Yau 3-fold:}
In this case, there are two fixed points, $f=N,S$ the
North and the South poles of the sphere, the $\Omega$ field descends from
the generic toric rotation of ${\bC}^4$ of which the local ${\bP}^1$ is the
quotient:
$$
(z_0, z_1, z_2, z_3) \mapsto ( {\l} t_0 z_0 , {\l} t_1 z_1, {\l}^{-1}t_2 z_2, {\l}^{-1}t_3
z_3), \qquad {\l}, t_{i} \in {\bC}^*
$$
The action of $s\in {\bC}^{*}$ on $z_{0,1,2,3}$ follows from the linear sigma model
charges for this geometry given by $(1,1,-1,-1)$. The toric diagram of this geometry is
shown in \res.
The North pole is the point $(1:0:0:0)$ and the South
pole is $(0:1:0:0)$. The tangent space at the North pole is ${\bf C}^{3}$ parameterized by
$(z_{1},z_{2},z_{3})$ and the tangent space at the South pole is also ${\bf C}^{3}$ but
parameterized by $(z_{0},z_{2},z_{3})$. The local weights given by the eigenvalues of the toric action
on the tangent space are: at $N$: $({\e}_1 -
{\e}_0, {\e}_2 + {\e}_0 , {\e}_3 +{\e}_0)$, at $S$:
$({\e}_0 - {\e}_1, {\e}_2 + {\e}_1, {\e}_3+{\e}_1)$.

\ndt{\bf Non-generic ${\bP}^1$ in Calabi-Yau 3-fold:}
This is the total space of the bundle ${\CO}(-m_1) \oplus {\CO}(-m_2)$ over
${\bP}^1$, with $m_{1}+m_{2}=2$. The condition $m_{1}+m_{2}=2$ follows from the total
space being Calabi-Yau. The linear sigma model construction of this space has a single
$U(1)$ gauge group with the charge
vector: $(1,1,-m_{1}, -m_{2})$. This geometry can be described as a quotient of ${\bf C}^{4}$ by the following
action:
$$(z_{0},z_{1},z_{2},z_{3})\rightarrow (sz_{0},sz_{1},s^{-m_{1}}z_{2},s^{-m_{2}}z_{3})\ , \qquad s\in {\bC}^{*}\ .$$

\ifig\ress{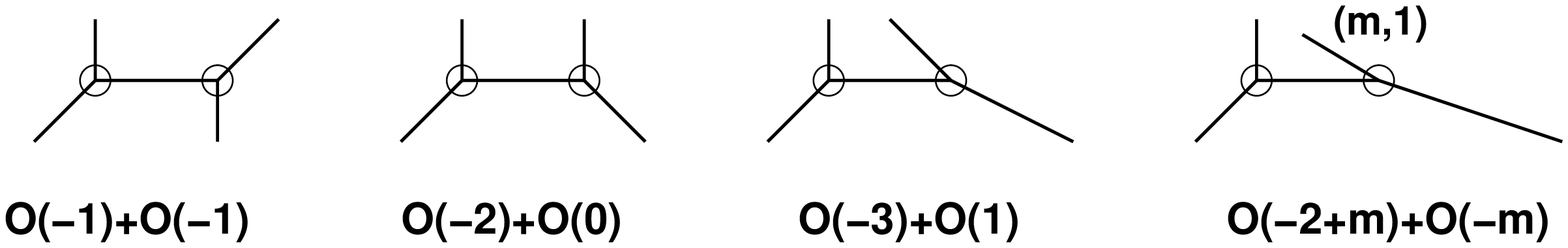}{110}
{The toric diagram of ${\cal O}(-2-m)\oplus {\cal O}(m)$ over ${\bf P}^{1}$. The two small circles
show the fixed points of the geometry.}
 The toric diagram is shown in \ress.

There are again two fixed points $f=N$ and $f=S$, with the local weights:
\eqn\lcwgmm{\eqalign{&( {\e}_{1N}, {\e}_{2N}, {\e}_{3N}) = ({\e}_1 - {\e}_0, {\e}_2 + m_1 {\e}_0, {\e}_3 + m_2 {\e}_0)\cr
& ( {\e}_{1S}, {\e}_{2S}, {\e}_{3S}) = ({\e}_0 - {\e}_1, {\e}_2 + m_1 {\e}_1, {\e}_3 + m_2
{\e}_1)\cr}}
respectively. Note that in both examples, if ${\e}_0 + {\e}_1 + {\e}_2 +
{\e}_3 = 0$ then ${\e}_{1f} + {\e}_{2f} + {\e}_{3f} = 0$ for all fixed
points. This is of course true for any Calabi-Yau $X$. Also, note that
\eqn\eminem{( {\e}_{1S}, {\e}_{2S}, {\e}_{3S}) = ( -{\e}_{1N}, {\e}_{2N}+m_1 {\e}_{1N}, {\e}_{3N}+m_2
{\e}_{1N})}

\ndt{\bf Weighted ${\bP}^2$ in Calabi-Yau 3-fold:} Let us consider the
case of ${\cal O}(m_{0})$ bundle
over the weighted projective space ${\bP}^{2}_{[m_{1},m_{2},m_{3}]}$ with weights $(m_{1},m_{2},m_{3})$.
\ifig\wpp{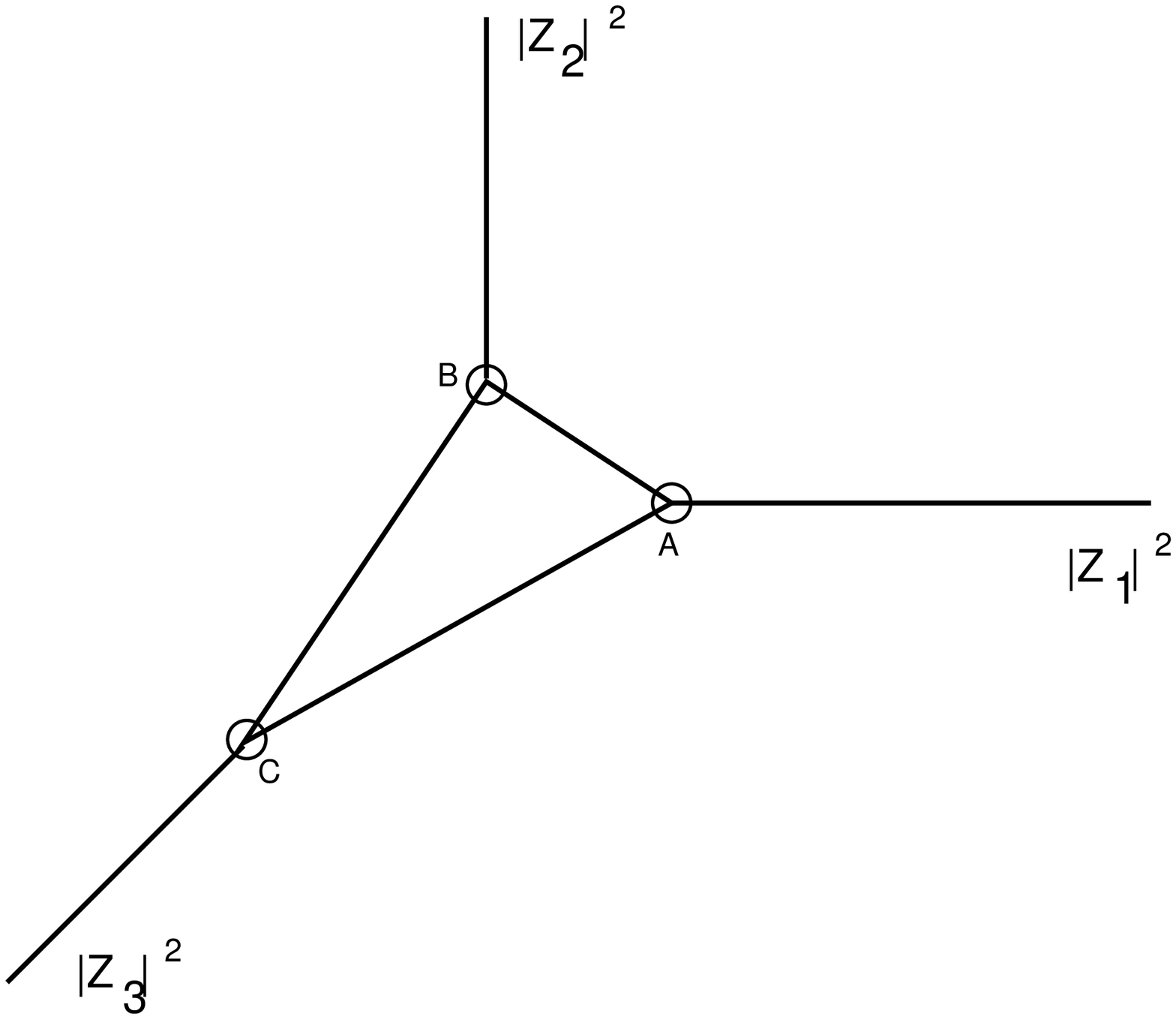}{60}
{Toric diagram of ${\cal O}(m_{0})$ bundle over the weighted projective space ${\bP}^{2}_{m_{1},m_{2},m_{3}}$. The fixed points of the
torus action are shown by small circles.}
The toric diagram is shown
in \wpp.  The
linear sigma model charges for this case are given by $(m_{0},m_{1},m_{2},m_{3})$,
\eqn\wei{(z_{0},z_{1},z_{2},z_{4})\mapsto (s^{m_{0}}t_{0}z_{0},s^{m_{1}}t_{1}z_{1},s^{m_{2}}t_{2}z_{2},s^{m_{3}}t_{3}z_{3}).}

This geometry has three fixed points $\{A,B,C\}$ are given by $(z_{0},z_{1},z_{2},z_{3})=(0,1,0,0),(0,0,1,0),(0,0,0,1)$ respectively.
These three points are not smooth but rather have orbifold singularities of order $m_{1},m_{2},m_{3}$.
The corresponding weights $\epsilon_{\alpha f}$ can be calculated using \wei,
$$(\epsilon_{1\ A},\epsilon_{2\ A},\epsilon_{3\ A})=(\epsilon_{0}-{m_{0}\over m_{1}}\epsilon_{1},\epsilon_{2}-{m_{2}\over m_{1}}\epsilon_{1},\epsilon_{3}-{m_{3}\over m_{1}}\epsilon_{1})$$
$$(\epsilon_{1\ B},\epsilon_{2\ B},\epsilon_{3\ B})=(\epsilon_{0}-{m_{0}\over m_{2}}\epsilon_{2},\epsilon_{1}-{m_{1}\over m_{2}}\epsilon_{2},\epsilon_{3}-{m_{3}\over m_{2}}\epsilon_{2})$$
$$(\epsilon_{1\ C},\epsilon_{2\ C},\epsilon_{3\ C})=(\epsilon_{0}-{m_{0}\over m_{3}}\epsilon_{3},\epsilon_{1}-{m_{1}\over m_{3}}\epsilon_{3},\epsilon_{2}-{m_{2}\over m_{3}}\epsilon_{3})$$

Note that the sum of weights at each fixed point are such that $\sum_{\alpha=1}^{3}\epsilon_{\alpha f}=0$ ($f=A,B,C$)
if $\epsilon_{0}+\epsilon_{1}+\epsilon_{2}+\epsilon_{3}=0$ and
$m_{0}+m_{1}+m_{2}+m_{3}=0$. The latter condition
is required for the total space to be a Calabi-Yau threefold.

Let us now return to the linear sigma model. The Calabi-Yau condition $\sum_i Q_i^a
= 0$ entails a natural constraint on the choice of twisted masses, or weights:
\eqn\cycnd{\mathboxit{\sum_{i=1}^{3+r} {\e}_i = 0}}
It follows that at each fixed point $f$ the local weights sum to
zero:
${\e}_{1f} + {\e}_{2f} + {\e}_{3f} = 0$.

\newsec{Application to ${\bf C}^3$}

In section 2 we proposed that the target space
description of topological A-model can be viewed
as sums over distinct K\"ahler manifolds, subject
to quantization of K\"ahler moduli in units of $g_s$.
In section 3 we gave examples of toric blow ups of ${\bf C}^3$.
In this section we propose that for toric geometries, such as
${\bf C}^3$, the sum over distinct K\"ahler manifolds localizes
on distinct toric geometries.
This we conjecture to be the case at least
as long as one is probing the geometry with toric branes.
A natural way this can be accomplished is proposed
in section 6.  As far as the discussion in this
section and section 5 is concerned we assume such
a localization holds with a simple measure and postpone
a technical justification of this assumption to section 6.
In fact we propose following \fundr\
$$Z=\sum_{quantized\ toric\ blow \ ups} {\rm exp}({Ng_s})\,,$$
where $N$ denotes the number of holomorphic sections
of the bundle whose curvature is $k/g_s$.

We can also turn
the question around: We ask which configuration of
sections, and
thus which values of $N$ are allowed?  If $|\psi \rangle \in {\cal H}$ is a
holomorphic section, then so is
$$z_1^{n_1} z_2^{n_2} z_3^{n_3} |\psi \rangle$$
for $n_i \geq 0$.  This is because we are not changing the complex structure
of ${\bf C}^3$ but only blowing up the geometry,
which modifies its K\"ahler geometry.  If we denote the
state
$|\psi \rangle =|m_1,m_2,m_3\rangle$, then as discussed before the
above action implies that
$$|m_1,m_2,m_3\rangle \in {\cal H}\rightarrow |m_1+n_1,m_2+n_2, m_3+n_3\rangle
\in {\cal H}$$
as long as $n_i \geq 0$.

The set of such ${\cal H}$  is in one to one correspondence
with 3D Young diagrams.  Counting the states of such ${\cal H}$
compared to that of ${\bf C}^3$
can be viewed
 as counting the negative of the
number of boxes of the 3D Young diagram.  Thus if we
restrict to this set we precisely obtain the partition associated with crystal
melting, but now viewed as summing
up over various K\"ahler geometries, obtained by
toric blowups of ${\bf C}^3$.  This is in fact
not exactly true, because we have not
shown that all K\"ahler geometries of
toric blow ups give rise to such ${\cal H}$.
Indeed this is {\it not} the case:
As we discussed before the points on the toric
base always form a convex set.
But this is not necessarily true for all ${\cal H}$.
For example, consider a 3d Young diagram
given by a cube of size $N$.  The complement
of the integral points on the octant
is not a convex set, thus it does not correspond
to a geometric blow up of ${\bf C}^3$.  It should be clear
from our discussion of toric geometry in section 2
that a 3D Young diagram corresponds
to a blow up geometry if and only if its complement is a convex set.
Given the precise match which was obtained
from the 3d Young diagram partition function and topological string in
\orv\ it is natural to postulate that we should extend the space of
geometries we are summing over to include such
cases of `non-geometric excitations'
of ${\bf C}^3$.  These `non-geometric' excitations
turn out to correspond to excitations of the gauge theory
discussed in section 6, which define the target space topological
gravity.
  That one may
need to sum over non-geometric excitations to obtain a
consistent quantum theory of gravity was already observed in three
dimensional gravity in \wittencsg\ where it was
seen that certain configurations of $SL(2,{\bf C})$ Chern-Simons
connection correspond to
negative values of $det(g)$.  This is probably not unrelated to the
convexity/concavity we have found here, which distinguishes geometric versus
non-geometric
excitations of ${\bf C}^3$: In fact the A-model K\"ahler gravity induces
on Lagrangian submanifolds the $SL(2,{\bf C})$
Chern-Simons gravity theory \guv.

Thus we see that with these assumptions, the sum over generalized
discrete toric geometries
contributes
$$Z=\sum_{\pi} q^{|\pi|}\,,$$
where $q={\rm exp} (-g_s)$, and $\pi$
denotes a 3D young diagram (or equivalently
a generalized toric geometry),
and $|\pi|$ the number of boxes in the Young diagram.
Thus we have mapped the sum over target K\"ahler geometries to the statistical
mechanics of crystal melting.  We now turn to the physical interpretation
of this result.

\subsec{Physical Interpretation of the Result}

We have thus mapped topological
A-model K\"ahler gravity to a 3D cubic crystal melting
problem, where the crystal occupies the positive octant. Moreover
the lattice spacing is $g_s$.  As the crystal melts we remove atoms, starting
from the corner.  The allowed configurations are such that the remaining atoms
form a crystal with arbitrary positive
translations in the three directions occupied
by an atom.  Each such configuration receives a thermodynamic weight
factor of $q^N$ where $N$ is the number of atoms removed from the crystal.
This statistical mechanical model has been studied in \cym .
The result is that for $g_s<<1$ (i.e. high
temperature for the crystal) the corner of the crystal gets molten up to a size
of order $1$ (in string units).

\ifig\ls{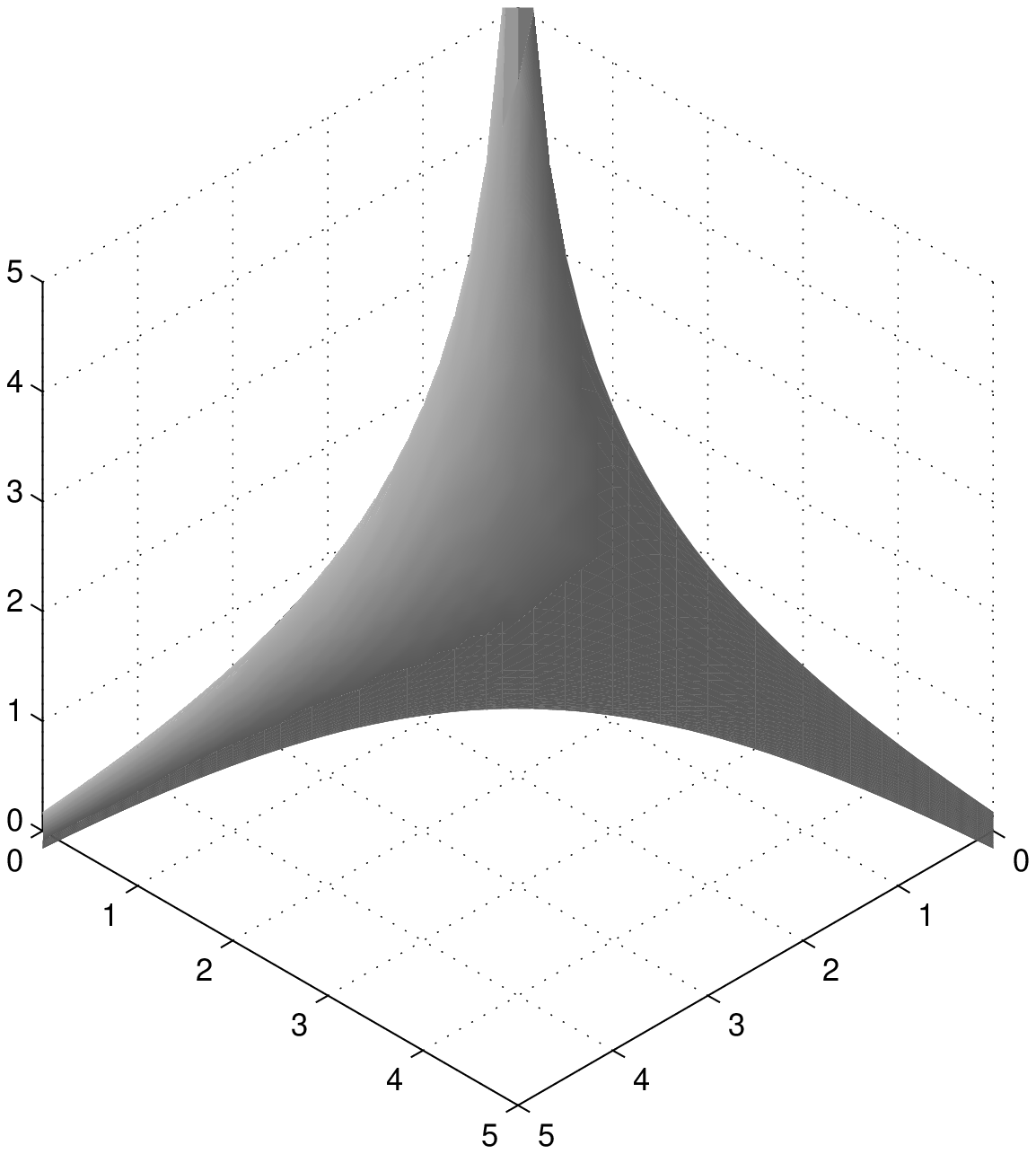}{60}
{The limit shape of 3d partitions.}

In particular the boundary of the crystal
becomes a smooth object at the string scale, known as
the limit shape (\ls), defined by
$$(p_1,p_2,p_3)=(u+R(u,v),v+R(u,v), R(u,v))$$
where
$$R(u,v)={1\over 4\pi^2}\int d\theta d\phi \
{\rm log}|e^{-u+i\theta}+e^{-v+i\phi}+1|$$
Note that the transition region of smooth (flat) and curved boundary of
the crystal
are defined by three curves intersecting
the three planes given by  $\pm e^{-u} \pm e^{-v}=1$ (excluding
the case where both terms have a minus sign).

\ifig\dpp{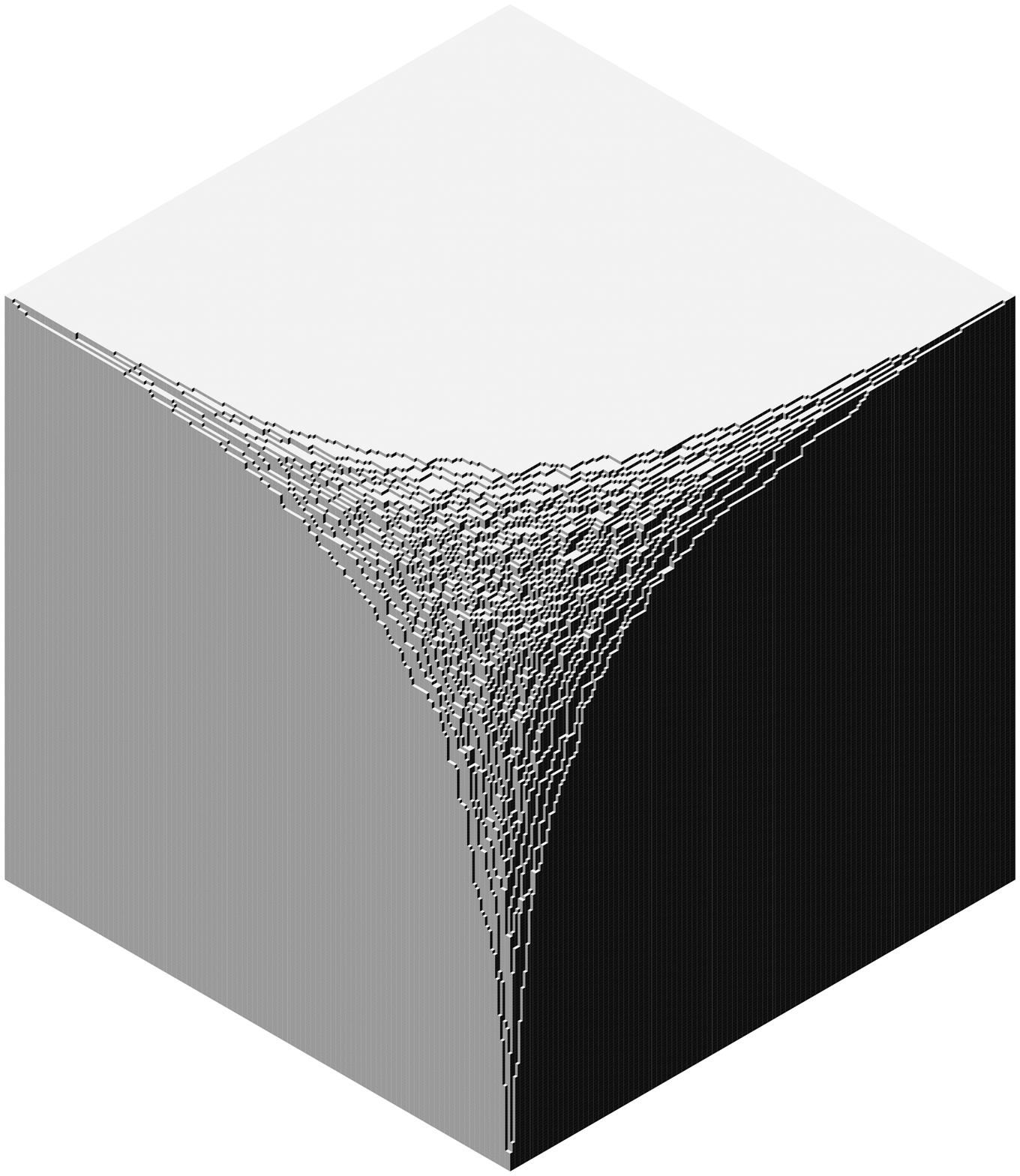}{90}
{A random partition with $g_{s}\approx 0.02$ inside a box of size
$200\times 200 \times 200$ lattice lengths.}

\ifig\dlss{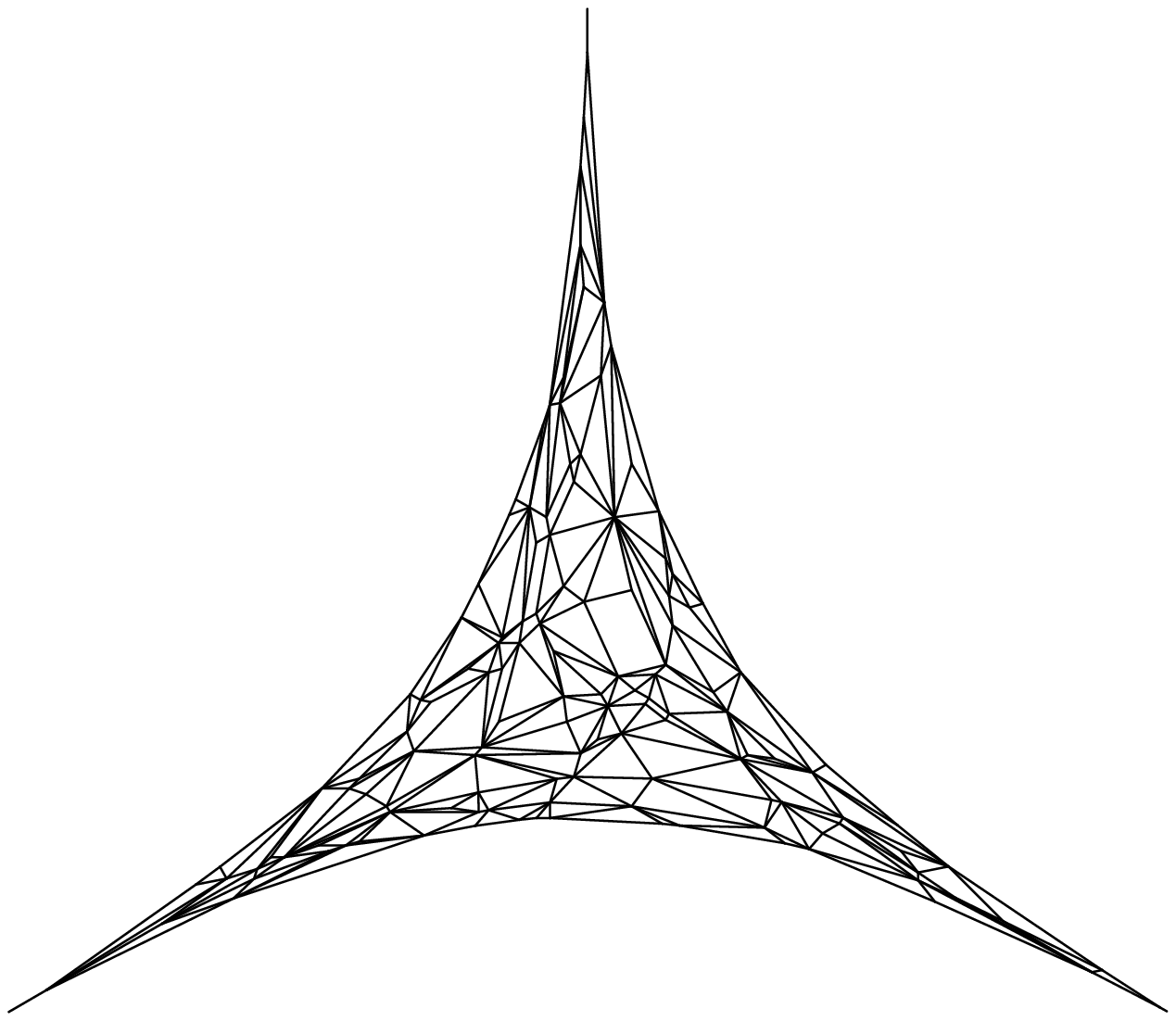}{70}
{The blownup geometry corresponding the 3d partition shown in \dpp .}

 The limit shape is the boundary of the base of ${\bf C}^3$ in the limit of very large number of
blowups. It is
essentially the region which replaces the origin in ${\bf
C}^{3}$,
in the limit where $g_s\rightarrow 0$ where the number of blowups
approach infinity, connected
to the rest of the flat planes of the octant. For a
finite but small $g_s$ there are a finite but large number
of faces of the blownup geometry, and the boundary is a piecewise linear
version of the limit shape made up of
polygons. The limit shape has the
same overall size, of the order of string scale, independent
of $g_s$.
These various polygons are the
compact four cycles in the blowup geometry.
The convex hull of the points closest to the limit shape
is the piecewise linear geometry we mentioned before.
The points also define a 3d partition.
The \dpp\ shows the 3d partition and \dlss\ shows
the corresponding blownup geometry for $g_{s}\approx  0.02$.

{}From the viewpoint of topological string we should identify the region
occupied
by the lattice as the toric base of the geometry.  The fact that the boundary
is deformed from the naive classical picture, is consistent with the fact that
worldsheet instantons modify the A-model target geometry at
string scale, captured
by a classical mirror geometry.
In fact, as noted in \orv\ the boundary of the molten crystal can be viewed
as a special lagrangian submanifold, which is captured by the mirror geometry
\refs{\hv , \hiv}\ encoded by the Calabi-Yau hypersurface
$$e^{-u}+e^{-v}+1=zw$$
The points of the blown up geometry are now deleted from space.  Thus
the molten piece of crystal also represents the ``molten space''.

Let us probe this geometry by toric Lagrangian A-branes.  As discussed
in
section 2, we consider non-compact A-branes which project
to an infinite diagonal semi-line intersecting the $p_1$ axis at the point
$a$.  Note that we are working
in the string scale and $a$ has units of area.  In other
words in the formulae below by $a$ we mean the dimensionless
quantity $a/l_s^2=a/\alpha'$.

\ifig\per{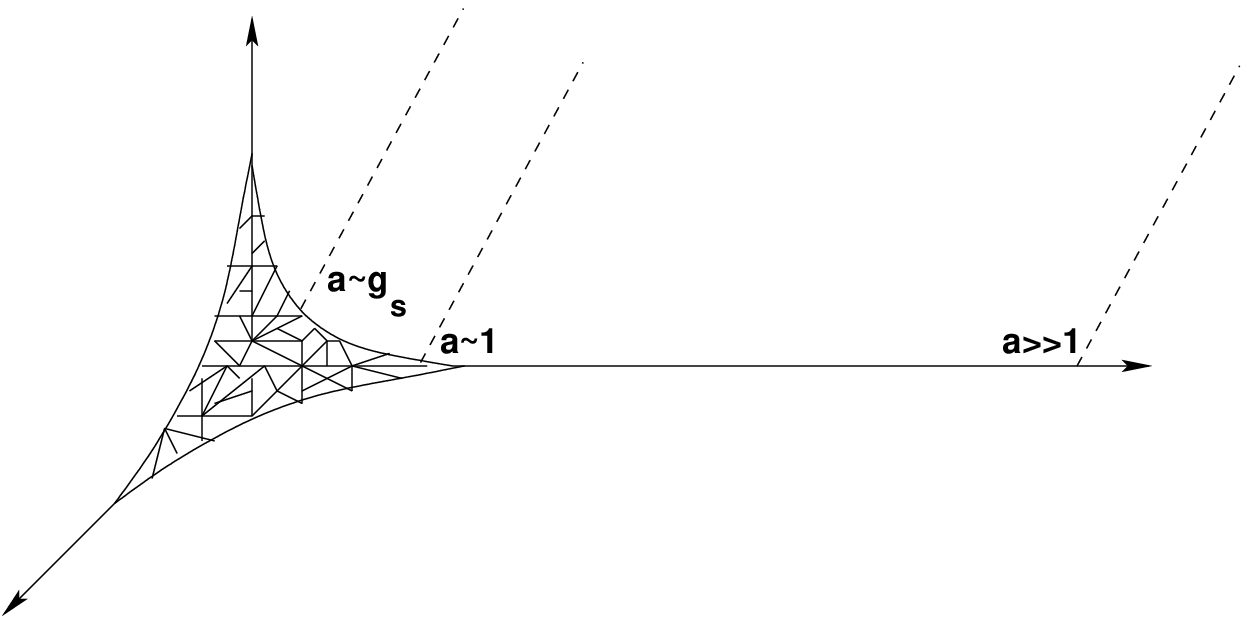}{90}
{As the A-brane probe moves there are three
regimes of geometry. For $a>>1$ it is classical.
For $a\sim 1$ it is stringy geometry and for $a\sim g_s$
it is a foamy geometry.  The tree level worldsheet instanton
 corrections can be read off from the geometry of the three curves
bounding the curved portion of the limit shape.}

We have three distinct regions for this probe (see \per ):
$$a>>1\qquad classical \ regime$$
$$a\sim 1 \qquad stringy \ regime$$
$$a\sim g_s \qquad quantum \ foam \ regime $$

For $a>>1$ the melting of the crystal is irrelevant and
we just see the full octant.  This is consistent with the fact
that in this region the worldsheet instantons are
highly suppressed $e^{-a}\sim 0$,
and we recover the classical geometry of ${\bf C}^3$.  For $a\sim 1$ we come
close to the boundary of the crystal and we see the limit
shape.  We see this as a smooth geometry at this scale.  This is
the Lagrangian boundary of the space.  In this regime the worldsheet
instantons become relevant and give rise to the stringy geometry
which is captured by the limit shape.  In this regime it is important
to note that the brane probe position is not strictly on the $p_1$ axis
as that has molten away.  Rather its end is replaced by the boundary regions
of the limit shape, say $e^{-u}+e^{-v}=1$, where
$$u=p_1-p_3$$
$$v=p_2-p_3$$
In other words in the classical regime the Lagrangian
A-brane defined by $u=a, v=0$ was denoting the moduli
of the brane.  In the quantum theory the moduli of brane is projected to
$u,v$ satisfying
$$e^{-u}+e^{-v}=1$$
Note that for $u=a>>0,v=0$ this is satisfied.  But for $a\sim 1$
the allowed $u,v$ change.  The leading string correction
to the partition function (i.e. the disc amplitude)
due to the presence of the brane was computed
from this curve \mic\ and is given by
$$F_0={-1\over g_s}\int^a v(u) du ={1\over g_s}\sum_{n>0}
 {e^{-na}\over n^2}.$$
Note that $e^{-a}$ represents the amplitude for the disc instanton ending
on the probe and $n$ labels the multicoverings of this instanton.
Note that for small $a$ this diverges as $F_0={1\over g_s} a{\rm log}a$.
One can see that as $a\rightarrow 0$ the string perturbative expansion
diverges.  In fact as shown in \ov\ the full amplitude
can be computed using the duality between Chern-Simons on $S^3$ and topological
strings on resolved conifold \gova\ and is given by
$$ F=\sum_{n>0} {e^{-na}\over 2n \ {\rm sinh}\left( {ng_{s}\over 2}\right)}$$
and expanding in powers of $g_s$ we find that for small $a$
$$F_g\sim c_g g_s^{2g-1}/a^{2g-1}$$
We thus see that as $a\sim g_s$ the string perturbation theory breaks down.
This is the regime where we see the quantum foam and the fluctuations of
the K\"ahler geometry. At this scale the geometry is fluctuating
rather wildly and the string expansion does break down.  Nevertheless
the exact expression continues to hold
(with a natural analytic continuation) even in this regime.  This
expression
can also be written in the more suggestive form \tvtwoo
\eqn\exe{Z=e^{F}=\prod_{n\geq 0}\left(1-e^{-a} q^{n+{\half}}\right)}
In fact, as explained in \tvtwoo\ this satisfies
a difference equation involving shifting $a\rightarrow a+g_s$:
\eqn\exac{Z(a+g_s,g_s)=\left(1-e^{-a} q^{\half}\right) Z(a,g_s)}
which is a reflection of the integral lattice structure at the
scale of $\delta a\sim g_s$.

Note that the foamy nature of spacetime takes place
at all points, and not just the origin of ${\bf C}^3$, as it
may appear from the discussion above.
Toric brane probes are sensitive only to the fluctuations
of geometry near the origin.  Origin of ${\bf C}^3$ is the only place
where the area of the disc which ends on them becomes
zero and the branes
become classically singular.  Of course by defining
the toric actions differently we get different toric branes
which probe the foamy structure of ${\bf C}^3$ at other
points.

So far our discussion has been in the context of $g_s<<1$.
However we can use the lattice picture to study the regime of large
$g_s>>1$.  This is the low temperature phase for our crystal.  At
this point none of the atoms of the crystal have molten and so the
geometry is just the classical geometry of the octant.  It may appear
surprising that in the extreme large string coupling constant the classical
geometry dominates--even the worldsheet instantons do not survive. To
make this more intuitive, note that even though the action seems
to allow arbitrary fluctuations when $g_s>>1$ (being given by $\int
k^3/g_s^2$),
what suppresses fluctuations is the quantization of $k$ in units of $g_s$:
$k=N g_s$.
Effectively what this means is that if we want to blow up a point
in ${\bf
C}^3$ we have
to replace it by a big ${\bf P}^2$ of the size $g_s$.  This should
be clearly suppressed.  This is also somewhat similar to what one sees
in superstring dualities where large string couplings lead to a classical
regime.  Note that the exact answer \exe\ is consistent with
all the corrections disappearing in this limit.  As $g_s\rightarrow
\infty$, we have $q\rightarrow 0$ and $Z\rightarrow 1$.

\newsec{More General Toric Geometries and the Topological Vertex}

In this section we will discuss some more general non-compact toric Calabi-Yau
threefolds.
For these more complicated geometries,
which may have compact 2-cycles and 4-cycles,  the general
idea,
discussed in section 2, of summing over toric blow ups
with quantized K\"ahler classes continues to hold.
The extra ingredient due to the presence of compact 2-cycles is
that we can, in addition to blowing up
points,  also blow up along the 2-cycles.
However, the restriction to toric blow ups
implies that the blown up 2-cycles must be invariant
under the torus action and hence must be rational curves joining the
fixed points.  From our discussion of toric geometries it is
clear that we can view them as obtained by trivalent vertices
connected by edges.
Thus the choices  of the  blow up geometries along edges
localizes to
the choice of a 2d Young diagram $\mu_i$ for each edge.
If we associate to each edge a K\"ahler class $t_i = N_i g_s$ with
$Q_i=e^{-t_i}$, as discussed in section 2 we will get a factor
$$\prod_i Q_i^{|\mu_i|}$$
where $|\mu_i|$ is the number of points associated to the Young
diagram $\mu_i$.  Note that $N_i |\mu_i|$ are the number of
points being deleted from the integral points of the toric
base when we blow up the $i$-th edge along the
Young diagram $\mu_i$ and
$$e^{-g_s N_i |\mu_i|}=Q_i^{|\mu_i|}$$
represents the deficit contribution of the removal of these
points to the partition function.  Note that this formula
can also be viewed as $Q_i^{ch_2(\mu_i)}$, consistent with
our discussion of section 2.  On top of this the toric geometry
may have excitation along the vertices.  These get
weighted with $q^{ch_3}$ for each vertex.  Note that
for each trivalent vertex these excitations are on
top of the choices of the blow ups along the three
edges ending on the vertex, which are labelled
by three 2d Young diagrams $\mu ,\nu ,\lambda$.
The allowed excitation of the bundle with these fixed
choices gets identified with the crystal partition
function with three asymptotes $\mu, \nu , \lambda$ and
this leads to the topological vertex $C_{\mu ,\nu , \lambda}$ \tvtwo\
as shown in \orv .   This is again, assuming as before,
that we enlarge the space of geometries to include non-geometric
excitations.
 The technical meaning of this assumption
from the viewpoint of the $U(1)$ gauge theory instantons will
be discussed in section 6.  Note that
we get one factor of topological vertex for
each vertex.  Together with the edge contribution discussed above,
this reproduces the topological vertex sum rule
of \tvtwo\ for computation of A-model topological amplitudes
on toric CY  3-folds.

There are two subtleties in this.  One is the framing
factor.  This has been carefully considered in \orv\ and
it just is there to ensure the correct number of points
are counted in the partition sum, coming from the fact
that different corners of the toric geometry are oriented
differently.  In addition there is some $\pm$ sign factors
in the gluing rule of \tvtwo .
This can be absorbed into the definition of what $Q_i$ are
(see also our discussion in section 6 for the meaning
of these sign factors).

As an example consider
${\cal O}(-1)\oplus {\cal O}(-1)\rightarrow
{\bf P}^{1}$:
The toric diagram of this geometry is shown in \res\ .
As discussed in section 2
the partition function in this case can be written as
$$Z=\sum e^{{\cal S}}=\sum_{line \,bundles\, on\, toric\,
 blow ups}q^{ch_{3}}\,Q^{ch_{2}}$$
Where $t=-{\rm log}(Q)$ is the K\"ahler parameter of
the ${\bf P}^{1}$ and $ch_{2},ch_{3}$ are
Chern characters of the $U(1)$ bundle on (blown up) $X$.
Note that the term involving $ch_{2}$
was absent in the ${\bf C}^{3}$ case; this
was due to the fact that ${\bf C}^{3}$ had no compact 2-cycles.

To evaluate the above partition function note that in the case of
${\bf C}^{3}$ the contribution to $ch_{3}$ were in one to one
correspondence with 3D Young diagrams. In this case also we have
a correspondence between $ch_{3}$ and the 3D Young diagrams but in this
case $ch_{3}$ is determined by a pair of 3D Young
diagrams $(\pi_{1},\pi_{2})$
such that $ch_{3}=|\pi_{1}|+|\pi_{2}|$. This is due to the fact that there are
two fixed points under the $U(1)^3$ action.
In the melting crystal picture this corresponds to
removing integral points from both corners of this crystal.
If we take $Q\mapsto 0$ we essentially
get two copies of ${\bf C}^{3}$ and the partition function is the square of the
${\bf C}^{3}$ partition function,
$$Z(Q\mapsto 0)=\sum_{\pi_{1},\pi_{2}}q^{|\pi_{1}|+|\pi_{2}|}.$$
However, for non zero $Q$ the crystal
has an edge also connecting the two corners. The
integral points along this edge can also be
removed corresponding to blowing up the entire
${\bf P}^{1}$.
This has the effect of replacing the neighbourhood of ${\bf P}^{1}$
with ${\bf P}^{1}\times (Blow\ up \ of \ {\bf C}^2)$.
\ifig\ponefig{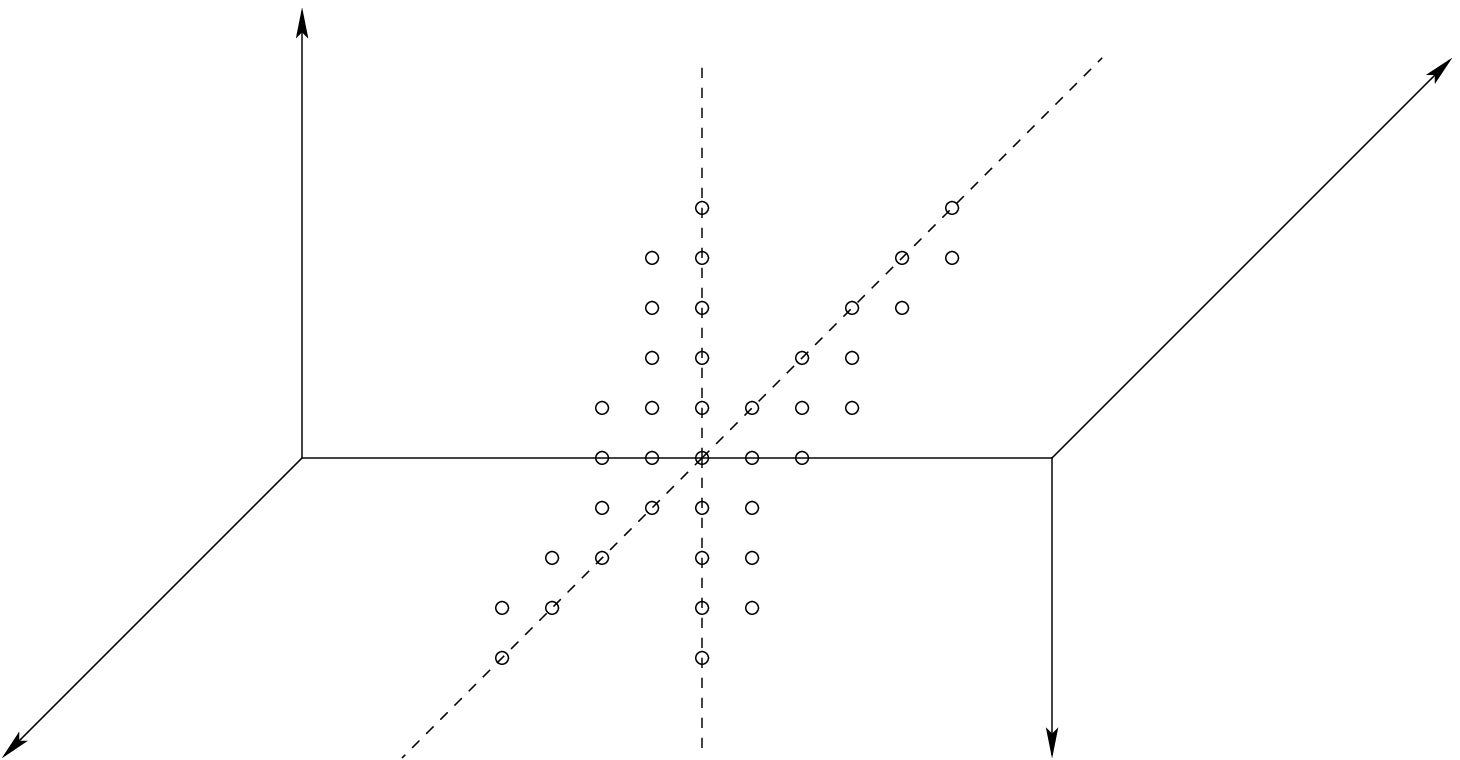}{90}
{More precisely the 2D partitions correspond to blowing up the entire
${\bf P}^{1}$.}
It is easy to see
that blowing up the entire ${\bf P}^{1}$ a
number of times is in one to one correspondence with
2D Young diagrams whose complement is convex \ponefig.
 Thus as in the case of ${\bf C}^{3}$ we not only have to consider
non-geometric excitation corresponding to
3D Young diagrams whose complement is not convex, but also have to
consider the non-geometric excitations
corresponding to 2D Young diagrams whose complement is concave.
Thus the contribution to $ch_{2}$ is in one to
one correspondence with 2D Young diagrams
essentially because the ${\bf P}^{1}$
is the fixed point locus of the $U(1)^2$ action.
Since the vertices are also invariant under
this $U(1)^2$ action the 2D Young diagram
must be compatible with the 3D Young diagrams coming from the vertices.
This is also reflected by the fact that
it is not possible to blow up the entire ${\bf P}^{1}$ without
blowing up the corners as well. This implies
that the 2D Young diagram, $\lambda$, must be
one of the asymptotes of the 3D Young diagrams
in the direction given by the ${\bf P}^{1}$.
Thus the partition function becomes
$$Z_{X}=\sum_{\lambda,\pi_{1},\pi_{2}}q^{|\pi_{1}|+|\pi_{2}|+(||\lambda||^2+||\lambda^{t}||^2)/2}Q^{|\lambda|},$$
$$=\sum_{\lambda}Q^{|\lambda|}\sum_{\pi_{1},\pi_{2}}q^{|\pi_{1}|+|\pi_{2}|+(||\lambda||^2+||\lambda^{t}||^2)/2}.$$
Where the factor involving $||\lambda||^2 \ (=\sum_{i}\lambda_{i}^{2})$
in the above equation is due to counting of
extra points arising due to perpendicular slicing
(see \orv) of the 3D Young diagrams associated with the
two vertices, \ponefig. As shown in \ponefig\ the slicings can be rotated
into each other by first rotating one of them
along the first column of the 2d partition $\lambda$ by $\pi/4$ and
then rotating it along the first row of $\lambda$ by
$\pi/4$. Rotation around the column gives an extra volume ${||\lambda||^2 \over 2}$
and rotation around the row gives an extra volume ${||\lambda^{t}||^2 \over 2}$.
The sum over 3D Young diagrams $(\pi_{1},\pi_{2})$ in the above equation is restricted by the
condition that their asymptotics in the direction of ${\bf P}^{1}$ is $\lambda$.
\ifig\per{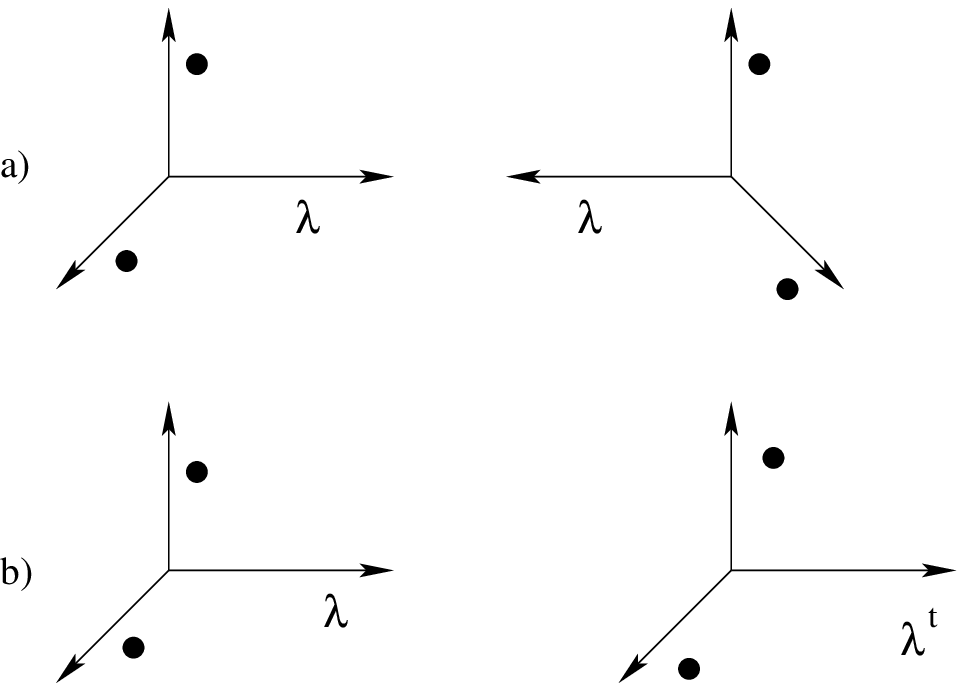}{90}
{a) The 3D Young diagrams associated with the vertices have
asymptote $\lambda$ along the ${\bf P}^{1}$, b)
in terms of the topological vertex this implies a gluing rule
which involves taking the transpose.}

This sum can be evaluated in terms of the topological vertex,
 $C_{\mu,\nu,\lambda}$, using the results of
\orv\ (see \per) ,
$$\sum_{\pi_{1},\pi_{2}}q^{|\pi_{1}|+|\pi_{2}|+(||\lambda||^2+||\lambda^{t}||^2)/2}=M(q)^{2}\ C_{\bullet \bullet \lambda}(q^{-1})C_{\bullet  \bullet \lambda^{t}}(q^{-1}).$$
Where $M(q)=\prod_{n>0}(1-q^{n})^{-n}$ and $\bullet$ is the empty partition.
Thus the partition function is given by
$$Z=M(q)^{2}\sum_{\lambda}Q^{|\lambda|}(-1)^{|\lambda|}C_{\bullet \bullet \lambda}(q^{-1})C_{\bullet \bullet \lambda^{t}}(q^{-1}),$$
$$ Z=M(q)^2\ \prod_{k>0}(1-Q\ q^{k})^{k}\ .$$
Where we have shifted $Q\rightarrow -Q$ in order to agree with the gluing rules of the topological vertex.

\ndt{\bf $T^{*}({\bf P}^{1})\times \bC$:} This geometry can be obtained from local ${\bP}^{1}\times {\bP}^{1}$
in the limit when the area of one of the ${\bP}^{1}$'s is taken to infinity. It will be useful to discuss this example
because, unlike the previous example, this is the simplest geometry that involves non-trivial framing. The linear sigma model
charges for $T^{*}({\bP}^{1})$ are given by $(1,1,-2)$, the field describing $\bC$ is, however, uncharged under this $U(1)$,
$$p_{1}+p_{2}-2p_{0}=t\ ,\quad p_{3} \geq 0\ .$$
\ifig\cot{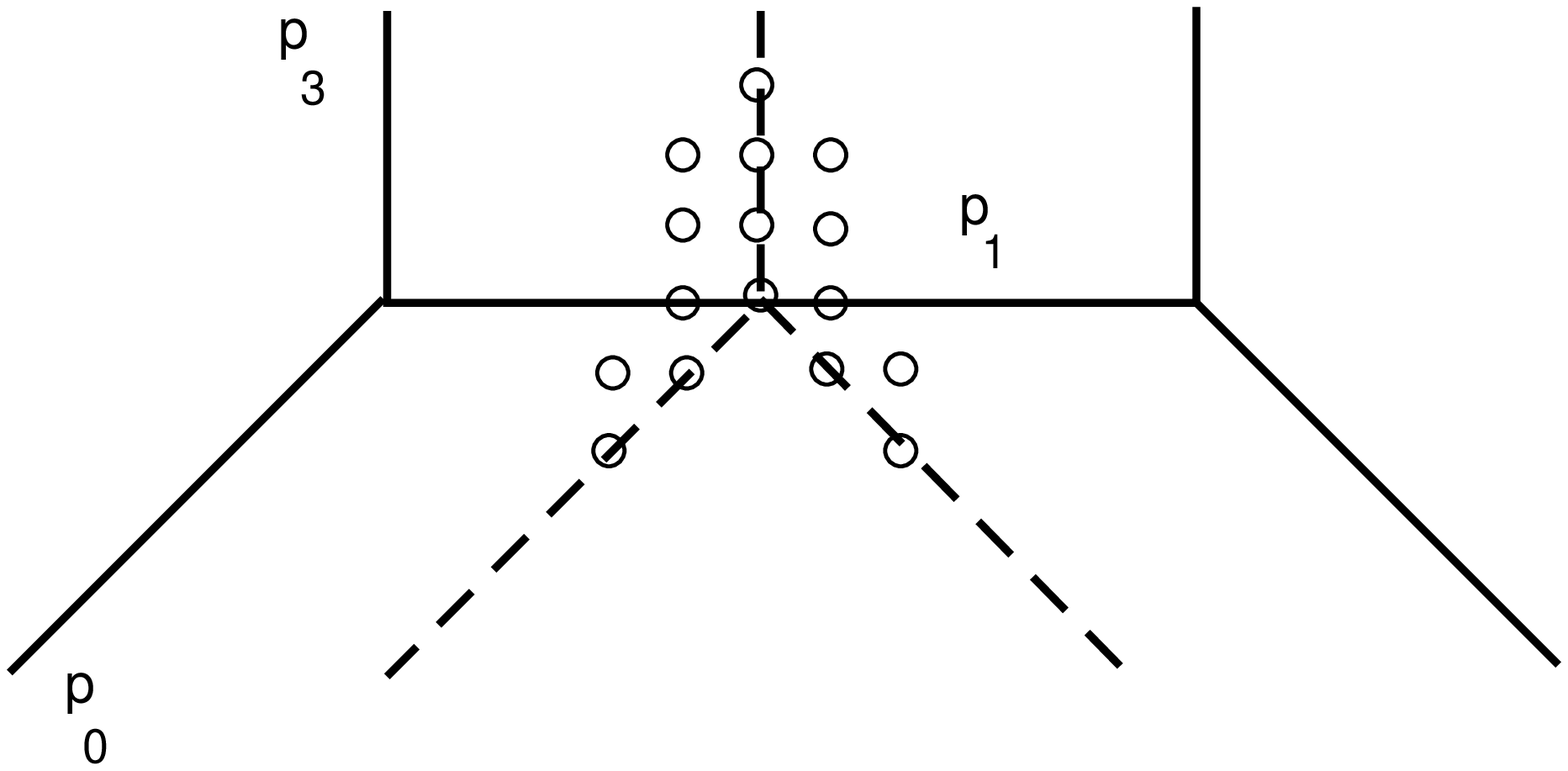}{90}
{Toric diagram of $T^{*}({\bf P}^{1})\times \bC$. Dashed lines indicate the perpendicular slicing of the
of the 3d partitions leading to 2d partitions on the dashed plane.}
\cot\ shows the toric diagram of this geometry. As discussed previously, the partition function is given by
$$Z=\sum_{\pi_{1},\pi_{2}}q^{|\pi_{1}|+|\pi_{2}|+\Delta(\lambda)|}\ Q^{|\lambda|}\ =\sum_{\lambda}Q^{\lambda}\sum_{\pi_{1,2}}q^{|\pi_{1}|+|\pi_{2}|+\Delta(\lambda)}.$$
Where the sum is over two 3d partitions $\pi_{1},\pi_{2}$ such that the 2d partition obtained by slicing them
in the direction of the ${\bP}^{1}$ is $\lambda$, and $\Delta(\lambda)$ is the volume between the two perpendicular slices
shown in \cot\ . If the 2d partition $\lambda$ is rotated on its first column by $\pi/4$ the volume obtained is given by
${1 \over 2}\sum_{i}\lambda_{i}^2={||\lambda||^2\over 2}$. This was the case in the previous example. In the example we
are considering now the rotation is such that it maps $(0,1)$ vector in the $(p_{1},p_{0})$ plane to $(2,1)$. Thus $\Delta(\lambda)=\sum_{i}\lambda_{i}^{2}=||\lambda||^2$.
 Using the definition of the topological vertex in terms of 3d partitions \orv\ we get
$$Z=M(q)^2\sum_{\lambda}Q^{\lambda}C_{\bullet \bullet \lambda}(q^{-1})C_{\bullet \bullet \lambda^{t}}(q^{-1})\ q^{\Delta(\lambda)-(||\lambda||^2+||\lambda^{t}||^2)/2}$$
$$=M(q)^2\sum_{\lambda}Q^{\lambda}C_{\bullet \bullet \lambda}(q^{-1})C_{\bullet \bullet \lambda^{t}}(q^{-1})\ q^{(||\lambda||^2-||\lambda^{t}||^2)/2}.$$
Using the relation $||\lambda||^2-||\lambda^{t}||^2=\kappa(\lambda)=|\lambda|+\sum_{i}\lambda_{i}(\lambda_{i}-2i)$ the partition function becomes
$$Z=M(q)^2\sum_{\lambda}Q^{|\lambda|}C_{\bullet \bullet \lambda}(q^{-1})C_{\bullet \bullet \lambda^{t}}(q^{-1})\ q^{\kappa(\lambda)/2}$$
$$=M(q)^2\sum_{\lambda}Q^{|\lambda|}C_{\bullet \bullet \lambda}(q) \ C_{\bullet \bullet \lambda}(q).$$
Where in the last equation we used the property of the Schur function $s_{\lambda^{t}}(q^{-i+1/2})=s_{\lambda}(q^{-i+1/2})q^{-\kappa(\lambda)/2}$.
Thus we see that correct framing factors follow from our prescription if points on the integral lattice are counted carefully.

\newsec{Target space theory viewpoint}

The aim of this section is to propose a precise (gauge) theory
which plays the role of target space K\"ahler gravity.
Just like the three (and two) dimensional gravities have gauge
formulations (involving $SL_{2}, E(2)$, etc.), K\"ahler gravity in any
dimension is in
some sense a $U(1)$ gauge theory.
In this section we first motivate the choice of the $U(1)$ gauge
theory as a twisted version of the maximally supersymmetric
theory in 6 dimensions. We then discuss how toric symmetry
localizes the path-integral for such a theory on certain toric
invariant subspaces.  The objects of relevance in the gauge theory
are 6d instantons, which are a singular configuration, very much
like the point-like $U(1)$ instantons in 4 dimensions.  To
make the point-like instantons more well defined\foot{It
is also interesting to pursue the other possibilities
of making the $U(1)$ instantons well defined, which came
up in the matrix model approach to ${\cal N}=1$ supersymmetric
gauge theory \refs{\dijky , \warny}:  Either considering $U(N)$ theory
and taking the $N\rightarrow 1$ limit, or more fundamentally
replacing $U(1)$ by the supergroup $U(k+1|k)$.}
 we also study a
certain
non-commutative
deformation of the $U(1)$ gauge theory, just as in the
4 dimensional case \nekin .

To motivate the following construction let us recapitulate what we
have said so far.
We want to study K\"ahler gravity on the space $X$. The space $X$ is a
noncompact (in most of what follows) toric Calabi-Yau manifold. It comes
with some base Calabi-Yau metric and the corresponding K\"ahler form $k_0$.
We want to study all K\"ahler metrics which asymptotically approach $k_0$.
We can write:
\eqn\klrfms{k = k_0 + g_s \ F}
where
\eqn\ffst{F^{2,0} = 0, \qquad d F = 0}
Here the $(2,0)$ projection is performed in the fixed complex structure of $X$ (of course, we only need the
almost complex structure, in agreement with the general philosophy of type A topological strings). The putative K\"ahler gravity equations of motion impose some constraints
on $F^{1,1}$. In quantum theory we calculate a path integral over
all possible $F$'s, as well as the corresponding ghosts, auxiliary fields and so on, with some supermeasure,
hopefully derived from the BV
formalism of \besa. Apparently, this superintegral has a $Q$-symmetry,
which according to the old Witten's argument localizes it onto the $Q$-fixed
points, which are the special solutions of the K\"ahler gravity equations of
motion.

Now note that the problem is reformulated in terms of the closed two-form,
which is of $(1,1)$ type and in addition satisfies some extra equation.
Imagine in addition that $F$ is quantized, i.e. its periods are
${\bZ}$-valued. Then we are dealing with $U(1)$ gauge theory on $X$.
We shall loosely call the $Q$-invariant solutions of this theory instantons.

\subsec{Supersymmetric Yang-Mills theory on six-manifold}

What gauge theory should we look at? If we embed the topological string into the superstring, as in \bcov\ then
the most natural choice is the
maximally supersymmetric gauge theory, i.e. the reduction
(with a twist) of
the ${\CN}=1$
super-Yang-Mills from ten to six dimensions. This theory was rediscovered in
many forms in the context of topological gauge theories in
\refs{\bks , \thom , \bsftop}.
In general one can study this theory for any gauge group $G$. The $Q$-fixed
field configurations there obey:
\eqn\ssyfc{F_{A}^{2,0} = {\pb}^{\dagger}_{A} {\varphi}}
\eqn\ssyfr{F_{A}^{1,1} \wedge k_0 \wedge k_0 + [ {\varphi}, {\bar\varphi} ]
= {\ell} \ k_0 \wedge k_0 \wedge k_0 ,}
where $\ell$ is some constant, fixed by the magnetic flux of the gauge bundle.
Here ${\varphi}$ is the adjoint-valued $(3,0)$ form, which is the twisted
complex scalar of the super-Yang-Mills theory. The second complex scalar
$\phi$
remains untwisted and is the analogue of the adjoint Higgs  field of the
four dimensional ${\CN}=2$ theory.
It must be covariantly constant:
\eqn\hgs{d_{A} {\phi} = 0}
The solutions of the system \ssyfc\ssyfr\ ought to be isolated,
the corresponding index being zero, but in practice this may not be so.

{}The bosonic part of the Lagrangian of our gauge theory is:
\eqn\bsnc{\eqalign{ L_{b} = &  {\half} {\tr} \left( d_{A}{\phi} \wedge \star d_{A}{\bar\phi} +
[{\phi}, {\bar\phi}]^2 + F_{A} \wedge \star F_{A} + D_{A}
{\varphi}\wedge\star
D_{A} {\bar\varphi} \right)\cr
& \qquad\qquad + \ {{\vartheta}\over{3! (2\pi)^3}} {\tr}
F_{A} \wedge F_{A} \wedge F_{A} \cr}}
the last line being the usual six dimensional theta term.
The fermionic content of the theory is the following: a one-form ${\psi}$,
two-form ${\chi}^{2,0}$,${\chi}^{0,2}$, three form ${\psi}^{3,0}$,${\psi}^{0,3}$, and two scalars ${\eta}$, ${\bar\eta}$
-- sixteen real components in total, as it should been.
The fermionic part of our Lagrangian is:
\eqn\fnnc{\eqalign{ L_{f} = & {\tr} \left( k_0 \wedge  {\chi}^{2,0}
\left( {\pb}_{A}{\psi}^{0,1} + {\pb}_{A}^{\dagger}
{\psi}^{0,3} \right)
 +  {\eta} \left( g^{i{\jb}} D_{i} {\psi}_{\jb}^{0,1} + [{\bar\varphi}, {\psi}^{0,3}]  \right) +   c.c. \right) \cr
+ &{\tr} \left( \left( k_0 \wedge [{\chi}^{2,0}, {\chi}^{0,2}] + [{\eta},
{\bar\eta}]
\right) {\phi} + \left( [ {\psi}, \star {\psi}] + [ {\psi}^{3,0}, {\psi}^{0,3}] \right) {\bar\phi}\right) \cr}}
By rewriting the action \bsnc\ in the form:
\eqn\bscnc{\eqalign{& S = \int_{X}{\half} {\tr} \left( d_{A}{\phi} \wedge \star d_{A}{\bar\phi} +
[{\phi}, {\bar\phi}]^2 + \vert F_{A}^{2,0} + {\pb}_{A}^{\dagger}{\varphi} \vert^2 + \vert F^{1,1}_{k_0}
\vert^2\right)
+\cr & \qquad\qquad \qquad {\half}{\tr}
\left( F_{A} \wedge F_{A} \wedge k_0 + {\vartheta \over 3} F_{A} \wedge
F_{A} \wedge F_{A}\right)\cr}}
we see that the solutions of \ssyfc\ssyfr\ provide the minima of the Yang-Mills
action in the given topological class, with precisely  \clsactn\ as the
instanton action (up to an additive constant, equal to the volume
of $X$) where
we identify
$\vartheta =i g_s$ (see below for a more precise map).

It is a standard exercise in the instanton calculus to work out the measure
on the space of collective coordinates. Linearization of the equations
\ssyfc\ssyfr\ together with the gauge fixing condition define some elliptic
complex, and the measure is the ratio of the determinants of the corresponding elliptic
operators. Due to supersymmetry the determinants almost cancel, leaving
essentially the zero modes to be taken care of. The standard by now trick is
to lift these zero modes by working with the nonvanishing vacuum expectation
value of the Higgs field $\phi$ and also by turning on the so-called
$\Omega$-background (see below). Before we plunge into details, let us
discuss the zero modes.

The moduli space ${\CM}$ of solutions to
\ssyfc\ssyfr\ for complex $X$ is a complex variety as well. Its holomorphic tangent space at the point $(A,{\varphi})$
is the space of solutions of the linearized
equations \ssyfc\ up to linearized complexified gauge transformations, i.e. the cohomology
of the complex:
\eqn\tngntspc{\eqalign{& \qquad\qquad \qquad\qquad\qquad\qquad {\CT}_{(A, {\varphi})} {\CM} = H^{1}\cr
& \cr
& \qquad 0 \longrightarrow {\Omega}^{0,0}_{A} \longrightarrow {\Omega}^{0,1}_{A} \oplus {\Omega}^{0,3}_{A} \longrightarrow
{\Omega}^{0,2}_{A} \longrightarrow 0
\cr
& \cr
& {\eta} \in {\Omega}^{0,0}_{A} \mapsto {\pb}_{A} {\eta} \oplus [{\varphi},
{\eta}] \ \in {\Omega}^{0,1}_{A} \oplus {\Omega}^{0,3}_{A}\qquad \cr
& \qquad\qquad\qquad\quad {\d}{\bar A} \oplus {\d}{\varphi} \ \in \ {\Omega}^{0,1}_{A} \oplus {\Omega}^{0,3}_{A}
\mapsto {\pb}_{A} {\d}{\bar A} + {\pb}_{A}^{\dagger}{\d}{\varphi} \in
{\Omega}^{0,2}_{A} \cr}}
In addition, on top of ${\CM}$, one finds the so-called {\it obstruction}
bundle ${\CN}$, or the antighost bundle in the physical terms, which is spanned by
the zero modes of the conjugate operator, or the $H^2$ of the complex above (which one finds by writing out the fermionic kinetic term for the theory \clsactn):
\eqn\obst{{\p}_{A}^{\dagger} {\chi}^{2,0} + {\p}_{A}{\eta} = 0, \qquad
{\p}_{A} {\chi}^{2,0} + [ {\bar\varphi}, {\eta}]= 0}
Integrating over the fermionic zero modes along $\CN$ brings down the
Pfaffian of the curvature of $\CN$.
The instanton contribution comes out to be:
\eqn\instcontr{\int_{\CM} \ {\rm Pfaff} \left({\CR}_{\CN}\right) = Euler({\CN})}
Note that this is not the Euler characteristics of $\CM$, unlike the four
dimensional maximally supersymmetric case \nfour, because $\CN$ does
not, in
general, coincide with ${\CT}\CM$, although it has the same rank.

\subsec{Toric localization}

Now recall that the space $X$ has a torus ${\bT}^3$ worth of isometries.
Then one can deform the Lagrangian \clsactn, $L \to L_{\Omega}$ by placing the theory in the so-called
$\Omega$-background, which in practice changes the terms $\| d_A {\phi} \|^2+ \| [{\phi},{\bar\phi}] \|^2$
to
\eqn\dfrm{\| d_{A} {\phi} - \iota_{\Omega} F \|^2 + \| [{\phi}, {\bar\phi}] - L_{\bar \Omega} {\phi}+ L_{\Omega} {\bar\phi}\|^2}
where $\Omega$ and ${\bar \Omega}$ are two commuting vector fields on $X$ generating
isometries (we won't see any of $\bar\Omega$ in what follows). The path integral in gauge theory will now localize onto the
solutions of \ssyfc\ssyfr\ which in addition are $\Omega$-invariant, or, in other
terms, such that the infinitesimal rotation of $X$ can be compensated by
the infinitesimal gauge transformation:
\eqn\fxdp{d_{A}{\phi} = \iota_{\Omega} F_{A}}
The
energy density
${\tr}F_{A}\star F_{A}$ of the $\Omega$-invariant solution wants to
concentrate near the $\Omega$-fixed
submanifolds in $X$.
Typically, the six dimensional instanton wants to look like a small size four
dimensional instanton fibered over a two-sphere, which connects two
fixed points $f_1$ and $f_2$ (and $\varphi$ vanishes). This takes care of the $ch_2 \sim {\tr}
F_A \wedge F_A$ part of the topological charge. In addition, for nontrivial
$ch_3 \sim {\tr} F_A^3$ the energy density has sharp peaks at the fixed points $f$ themselves.

Let us label the point-like instanton configurations concentrated near a
point $f$ by ${\pi}_f$ (${\pi}_f$ will turn out to be a 3d
partition) and the instantons spread over the two-sphere connecting two fixed points $f_1$ and $f_2$
by
${\l}_{f_1 f_2}$ (later on this will be a 2d partition). If we look at the
same configuration in the opposite direction, we shall label the
corresponding instanton by ${\l}^{t}_{f_1 f_2} = {\l}_{f_{2} f_{1}}$.
\eqn\instdnst{\eqalign{& {\tr}\left( F_{A} \wedge F_{A} \right)\sim \sum_{f_1, f_2}
|{\l}_{f_{1} , f_{2}}| {\d}^{(4)}_{{\bP}^1_{f_1 f_2}} \cr
& {\tr} \left( F_A \wedge F_A \wedge F_A \right)\sim \sum_{f} |{\pi}_f|
{\d}^{(6)}_{f}\cr}}

Now let us scale the physical metric on $X$ by a large factor (using the
complex nature of $k_0$ we can keep the class of $k_0$ intact), and apply
field theory factorization (the fermionic symmetry preserved by $\Omega$-background guarantees that
the partition function does not change). The path integral will therefore split as the
integral over local ${\bR}^6$ patches near each $f$, with asymptotic
boundary conditions set by ${\l}_{ff'}$ for all $f'$ connected to $f$. In
turn, the latter will be the sum over ${\pi}_f$ of the instanton
contributions evaluated at ${\pi}_f$.
Let us denote it by:
\eqn\instcnt{
{\Gamma}_{\l\m\n}(q, {\Omega}) = \int_{{\scriptscriptstyle asymptotics} \ {\l\m\n}} DA D{\phi} D{\varphi}
D\psi D\chi D\eta \
e^{-\int_{{\bR}^6} L_{\Omega}}, }
where $q = e^{i {\vartheta}}$, and by asymptotics ${\l\m\n}$ we mean \foot{Recall that torus-invariant instantons in ${\bR}^{4}$ are in one to one correspondence with 2d Young diagrams with instanton charge being the number of boxes in the Young diagram \neksw.} :
\eqn\asymt{\eqalign{ & z_1 \to \infty , z_2, z_3 \ {\scriptstyle finite}: \  A \to {\l} \ {\scriptstyle instanton \ on \ }
{\bR}^4 \cr &
z_2 \to \infty , z_1, z_3 \ {\scriptstyle finite}: \  A \to {\m} \ {\scriptstyle instanton \ on\ }
{\bR}^4 \cr &
z_3 \to \infty , z_2, z_1 \ {\scriptstyle finite}: \  A \to {\n} \ {\scriptstyle instanton \ on \ }
{\bR}^4 \cr}}
Then the partition function on $X$ is given by:
\eqn\fllprt{Z_{X}(q, {\Omega}) = \sum_{{\l}_e} \prod_{f} {\Gamma}_{{\l}_{e_1}{\l}_{e_2}{\l}_{e_3}}(q, {\Omega}_f) \prod_{e} e^{-t_{e}
|{\l}_{e}|}}
Here ${\Omega}_f$ is the local expression for ${\Omega}$ near $f$, $e_1,
e_2, e_3$ are the $\Omega$-invariant complex lines, passing through $f$,
$|{\l}_{e}|$ is the four dimensional instanton action , and $t_e$ is the
K\"ahler class of the line $e$.
For local ${\bP}^1$ the formula \fllprt\ specializes to:
\eqn\lcpon{Z_{local \ {\bP}^1}(q, {\vec\e}) = \sum_{\l} e^{-t |{\l}|}
{\Gamma}_{{\bullet},{\bullet},{\l}}(q,{\e}_1 -
{\e}_0, {\e}_2 + {\e}_0 , {\e}_3 +{\e}_0){\Gamma}_{{\bullet},{\bullet},{\l}^{t}} (q,{\e}_0 - {\e}_1, {\e}_2 + {\e}_1,
{\e}_3+{\e}_1)}

\subsec{Gauge vertex}

\ndt We now wish to learn more about the vertex
${\Gamma}_{\l\m\n}(q, {\Omega})$. To this end we study more closely the
gauge theory.
\bigskip
\ndt{\bf Flat space revisited: noncommutative interpretation:}
\ndt In the case $G=U(1)$ there are hardly any nontrivial solutions to
\ssyfc\ssyfr.
There is, however, a simple way out of this, at least on $X = {\bR}^6$ or
its orbifolds. Replace $X$ by its  noncommutative deformation. Say, on
${\bR}^6$ we work with the algebra:
\eqn\heisnb{[x^m, x^n] = i {\theta}^{mn}, \qquad m,n=1, \ldots, 6}
where ${\theta}^{mn}$ is some constant anti-symmetric matrix, which we shall assume to be of maximal rank.
Moreover, by an orthogonal rotation we can bring ${\t}$ into the form:
\eqn\nrmlf{{\t} = \pmatrix{
0 & {\t}_1 & & & & \cr
-{\t}_1 & 0 & & & & \cr
& & 0  & {\t}_2  & & \cr
& &  -{\t}_2 & 0 & & \cr
& & & & 0  & {\t}_3 \cr
& & & & -{\t}_3 & 0   \cr}, \quad {\t}_{\a}  > 0}
It is convenient to work with the fields:
\eqn\bckgr{X^n = x^n + i {\t}^{nm} A_{m}(x)}
as opposed to the gauge fields $A_{m}$. The field strength is related to the commutator:
\eqn\fld{[X^{m}, X^{n}] - i {\t}^{mn} = {\t}^{mm'}{\t}^{nn'} F_{m'n'}}
and the equations \ssyfc\ssyfr\ are replaced by:
\eqn\ssyfnc{\eqalign{&
[Z^{\a}, Z^{\b} ] + {\ve}^{\a\b\g} [ Z^{\dagger}_{\g}, {\varphi} ] = 0, {\a}, {\b} =
1,2,3\cr
& [Z^{\a}, Z^{\dagger}_{\a} ] + [{\varphi}, {\varphi}^{\dagger}] = 3}}
where we have introduced
\eqn\lncmb{Z^{1} = {1\over{\sqrt{2{\t}_1}}} \left( X^1 +  i X^2 \right),
\quad etc.}

The equation $d_{A}{\phi}=0$ is replaced by $[Z^{\a} , {\Phi}] = 0$.

In the vacuum:
\eqn\vcmsl{Z^{\a} = a_{\a}, \ {\varphi} = 0, \ {\Phi} = {\phi} \cdot 1\qquad\qquad [
a_{\a}, a_{\b}^{\dagger} ] = {\d}_{\a\b}}
Here is an example of the nontrivial solution to \ssyfnc, found in
\kraus (for $m=1$ in \opennc):
\eqn\inst{Z^{\a} = S \ a_{\a} \ \left[ 1  - {m(m+1)(m+2)\over{N(N+1)(N+2)}}
\right]^{\half}\ S^{\dagger}}
where:
\eqn\nmb{N = a_1^{\dagger}a_1 + a_2^{\dagger}a_2 +a_3^{\dagger}a_3}and
$S$ is the so-called Murray-von Neumann partial isometry:
\eqn\pisom{SS^{\dagger} = 1, \quad S^{\dagger}S = 1 - \sum_{i+j+k < m}\vert
i,j,k\rangle\langle i,j,k\vert}
and we have used the standard oscillator representation of the algebra of
$a_{\a}, a_{\b}^{\dagger}$.
This solution is related to the blow up of the point on ${\bC}^3$ discussed
above, with the K\"ahler form of exceptional ${\bC\bP}^2$ being $m$ times
the standard Fubini-Study form.

\bigskip
\ndt{\bf Toric localization of noncommutative gauge theory:}
Now we deform our gauge theory further, by turning on the so-called
$\Omega$-background \refs{\neksw ,\no }, which utilizes the toric symmetry of ${\bR}^6$.
The equation \hgs\ gets deformed to:
\eqn\hgsep{[{\Phi}, Z^{\a} ] = {\e}_{\a} Z^{\a}}
where ${\e}_1, {\e}_2, {\e}_3$ are the parameters of the
$\Omega$-background, the angles of the infinitesimal ${\bT}^3$ rotation of
${\bR}^6$.
One can show that the solutions to \ssyfnc\hgsep\ have the following form:
\eqn\instsl{Z^{\a} = S {\Lambda}^{-\half}_{\vec n} a_{\a}
{\Lambda}^{\half}_{\vec n} S^{\dagger} , \qquad {\Phi} = S \left( {\e}_1
a_1^{\dagger} a_1 + {\e}_2 a_2^{\dagger}a_2 +{\e}_3 a_3^{\dagger}a_3 \right)
S^{\dagger} }
and the partial isometry identifies the Hilbert space ${\CH}$ of all states
of the triple of harmonic oscillators,
\eqn\hlb{{\CH} = {\bC}[ a_1^{\dagger}, a_2^{\dagger},a_3^{\dagger} ]
|0,0,0\rangle}
with its subspace:
\eqn\hlbi{{\CH}_{\CI} = {\CI}(a_1^{\dagger},a_2^{\dagger}, a_3^{\dagger})  | 0,0,0\rangle}
where ${\CI} \subset {\bC}[w_1, w_2, w_3]$ is an ideal in the ring of
polynomials, generated by monomials. In the example above the ideal in
question is spanned by
\eqn\idls{w_1^{i}w_2^{j}w_3^{k}, \qquad i + j + k \geq m}
Any such monomial-generated ideal ${\CI}$ defines the so-called three-dimensional partition (which is infinite for nontrivial asymptotics):
\eqn\tdprt{{\pi} = \{ \ (i,j,k) \ \vert i,j,k \geq 1, \
w_1^{i-1}w_2^{j-1}w_3^{k-1} {\in\kern -.12in\slash \kern .1in} \ {\CI} \}
}
The physical picture of the localization is the following: the instantons
(which are point-like, or fuzzy of the ``size'' $\sim \t$) tend to sit on top of
each other, near the origin in ${\bR}^6$, the fixed point of the spatial
rotation. They can also, as we discussed above, concentrate along the
coordinate axes, which are also invariant under rotation, and asymptote to
the ordinary four dimensional noncommutative instantons \neksch, invariant under
rotations, and labelled by the two-dimensional partitions.
\bigskip
\ndt{\bf Back to determinants:}
What is the contribution of such an instanton? We should expand the
super-Yang-Mills action and pick up the determinants of the bosonic and
fermionic fluctuations. The ratio comes out to be (cf. \ihiggs):
\eqn\ratdet{e^{i{\vartheta} ch_3} {{{\Det}({\rm ad}{\Phi}) {\Det}({\rm ad}{\Phi}+{\e}_1+{\e}_2){\Det}({\rm ad}{\Phi}+{\e}_3+{\e}_2){\Det}({\rm
ad}{\Phi}+{\e}_3+{\e}_1)}\over{{\Det}({\rm ad}{\Phi} + {\e}_1 + {\e}_2+ {\e}_3)
{\Det}({\rm ad}{\Phi}+{\e}_1){\Det}({\rm ad}{\Phi}+{\e}_2){\Det}({\rm ad}{\Phi}+{\e}_3)}}
}where ${\rm ad\Phi}$ denotes the operator $[{\Phi}, \cdot ]$, and one
should omit the zero modes (actually these are absent).

Perhaps a more suggestive form of the ratio of the determinants is given by
the proper time representation:
\eqn\prop{Z_{\CI} ({\e}_1, {\e}_2, {\e}_3) = {\exp} i {\vartheta} \left[ {\Tr}_{\CH}e^{t{\Phi}}\right]_{t^{0}} + \int_{0}^{\infty} {dt\over t} (1 - e^{t{\e}_1})(1-e^{t{\e}_2})(1-e^{t{\e}_3}) {\Tr}_{\CH} e^{t{\Phi}} {\Tr}_{\CH} e^{-t{\Phi}}}
At any rate, this is an explicit function of the ideal ${\CI}$ and can be
calculated. We can actually split it as a contribution of the vacuum
solution with given asymptotics, and then some finite number of instantons
on top of it.
For the vacuum:
\eqn\vccm{{\Tr}_{\CH} e^{t{\Phi}} = {1\over
(1 - e^{t{\e}_1})(1-e^{t{\e}_2})(1-e^{t{\e}_3})},}
for ideal ${\CI}$ corresponding to the partition ${\pi}$ (perhaps infinite):
\eqn\find{{\Tr}_{\CH} e^{t{\Phi}} = {1\over
(1 - e^{t{\e}_1})(1-e^{t{\e}_2})(1-e^{t{\e}_3})} - \sum_{(i,j,k) \in \pi} e^{t({\e}_1 (i-1) + {\e}_2 (j-1) + {\e}_3 (k-1))},}
and the vacuum with fixed asymptotics ${\l}_1, {\l}_2, {\l}_3$:
\eqn\vccmas{{\Tr}_{\CH} e^{t{\Phi}} = {1\over
(1 - e^{t{\e}_1})(1-e^{t{\e}_2})(1-e^{t{\e}_3})} - \left[ {R_{{\l}_3}(e^{t{\e}_1},
e^{t{\e}_2})\over{(1-e^{t{\e}_3})}} + (3 \to 2 \to 1) \right]}
where
$$
R_{\l}(x,y) = \sum_{(i,j) \in {\l}} x^{i-1} y^{j-1}, \qquad | {\l} | =
R_{\l}(1,1)
$$
The instanton weight \prop\ splits, as:
\eqn\propr{Z_{\CI}({\e}_1, {\e}_2, {\e}_3) = Z^{vac}_{\l\m\n}({\e}_1, {\e}_2,
{\e}_3) Z^{fin}_{\CI}({\e}_1, {\e}_2, {\e}_3)}
where the last factor is a product of the finite (and equal) number of
terms in the numerator and the denominator.
Finally, the most amazing property of the instanton weight \prop\ is that
for ${\e}_1 + {\e}_2 + {\e}_3 = 0$ the finite part
$Z^{fin}$  is equal to $(-1)^{[{\CI}]}$, with $[{\CI}]$ defined more
precisely below, which allows us to identify, ${\Gamma}_{\l\m\n}\vert_{CY}
\sim C_{\l\m\n}$.

\ndt{\bf Remark.} Similar localization of supersymmetric gauge theory (not the simplification of the $Z^{fin}$ factors)
on toric manifolds
holds in any number of dimensions. In particular, in four real dimensions
one gets an expression for Donaldson invariants on any toric surface in
terms of the instanton partition function on ${\bR}^4$ \nikloc. In the
case
of $X = \widehat{{\bC}^2}$, the blowup of a point on ${\bC}^2$ this was
essentially used in \ny.

The formula \ratdet\ is written for $X = {\bR}^6$. For
general toric $X$ it should be modified. The factorization formula \fllprt\
suggests that the determinants of the fluctuations about the instantons
localized near the fixed points in $X$ and the connecting them spheres
should factorize as well. In fact, by relating the determinant ratio with
the index of appropriate family of elliptic operators (like ${\pb}_{A}
\oplus {\pb}^{\dagger}_{A}$), and using for the latter the equivariant
Atiyah-Singer index theorem, which expresses it as an integral over $X$, and
then using the latter Duistermaat-Heckmann \DHf\ formula we shall get
precisely \fllprt.
Again, for the vector field $\Omega$ preserving the holomorphic $(3,0)$ form
the determinants cancel up to a well-defined sign factor, discussed below.

\bigskip
\ndt{\bf Vertex and framing:}
Let us discuss a little bit more the vertex and the instanton factors.
We wish to weigh the instantons with the factor:
\eqn\iw{
{\exp} - \left[ {1\over 8{\pi}^2} \int_{X} k_0 \wedge {\tr} F_{A}^2 + {{\vartheta} \over 48{\pi}^3} \int_{X} {\tr}
F^3_{A}\right]}
which we can calculate for the $\Omega$-invariant instantons using
localization:
\eqn\lcls{\eqalign{& - {1\over 8{\pi}^2} \int_{X} k_0 \wedge {\tr} F_{A}^2 = \sum_{f} {{H_f {\CE}_f^{(2)}}\over{{\e}_{1f}{\e}_{2f}{\e}_{3f}}}
\cr & \cr & {i\over 48{\pi}^3} \int_{X} {\tr}F_{A}^3 = \sum_{f} {{{\CE}_f^{(3)}}\over{{\e}_{1f}{\e}_{2f}{\e}_{3f}}}\cr}}
Here:
\eqn\ef{{\CE}_f (t) = 1 - (1 - e^{t {\e}_{1f}})(1-e^{t{\e}_{2f}}
)(1-e^{t{\e}_{3f}}) \sum_{(i,j,k) \in {\pi}_{f}} e^{ t ( {\e}_{1f} (i-1) +
{\e}_{2f}(j-1) + {\e}_{3f}(k-1))},}
${\CE}_{f}^{(n)}$ denotes the coefficient in front of $t^{n}$ in the
expansion of \ef\ near $t=0$, and
$H_f$ is the value of the Hamiltonian \ham\ at the fixed point $f$ (note that shifting
$H$ by constant won't affect \lcls\ for compact $X$). Also, another
application of the fixed point formula gives:
\eqn\kcl{t_{e} = {{H_{{f}_{1}}-H_{f_2}}\over{{\e}_{e}}}}
for $e = (f_1 , f_2)$.
The character ${\CE}_f$ also appears in the expression for the ratio of the
determinants:
\eqn\gnxmsr{Z_{\CI}({\vec \e}) = {\exp} \int^{\infty} {{dt}\over t}
\sum_f {{{\CE}_f (t) {\CE}_f (-t)}\over{(1 - e^{-t {\e}_{1f}})(1-e^{-t{\e}_{2f}}
)(1-e^{-t{\e}_{3f}})}}}
We now want to specialize our torus action to respect the Calabi-Yau
condition.  Now, by examining
\lcls\ef\ we see that the $ch_2$ part of the instanton action, when summed over $f$'s will reduce to
\eqn\chtp{- \sum_e t_e | {\l}_{e}| ,}
where we should use \kcl. The $ch_3$ part actually depends on one parameter,
which we can take to be $x = {\e}_{2f}/{\e}_{1f}$.
The local contribution to the $ch_3$ part of the instanton action is equal to:
\eqn\chtrp{\eqalign{ &  {{{\CE}_f^{(3)}}\over{{\e}_{1f}{\e}_{2f}{\e}_{3f}}} = | {\pi}_{f}
| + I_{\l\m\n}\cr
& I_{\l\m\n} = \left[  \sum_{(i,j) \in {\l}} (i-{\half}) x + (j-{\half}) (-1-x)  \right]
+\cr & \qquad\qquad
\left[  {1\over x} \sum_{(j,k) \in {\m}} (j-{\half}) ( - 1 - x) + (k-{\half})
\right]+\cr & \qquad\qquad\qquad
\left[  -{1\over{1+x}}\sum_{(k,i) \in {\n}} (k-{\half})  + (j-{\half}) x \right]\cr}}
and
\eqn\fint{|{\pi}_{f}|  = \left[ \sum_{(i,j,k) \in {\pi}_f, \ i,j,k \leq N}\
1\
\right]-(N+1) \ \left( \ |{\l}|+|{\m}|+|{\n}|\ \right), \qquad N \gg 0}
Local considerations do not
give a canonical value for $x$. This is related to the framing of the topological vertex. However, in summing over the edges the $x$
dependence of the instanton action essentially cancels. More precisely, each edge $e=(f_{1}, f_{2})$
comes with the weights $(m_{1e},m_{2e})$\foot{the order $(m_1, m_2)$ is correlated with the orientation of the edge, by the complex three-form on CY}
 of the $\Omega$ action on the normal bundle
to the corresponding sphere ${\bP}^1_{e}$. The local weights $({\e}_{1f}, {\e}_{2f}, {\e}_{3f})$ at two
poles of the sphere are related by:
\eqn\wghts{({\e}_{1f_{1}}, {\e}_{2f_{1}}, {\e}_{3f_{1}}) \mapsto ({\e}_{1f_{1}}+m_{1e}e_{3f_{1}}, {\e}_{2f_{1}}+m_{2e}{\e}_{3f_{1}},
-{\e}_{3f_{1}})= ({\e}_{1f_{2}}, {\e}_{2f_{2}}, {\e}_{3f_{2}})}
as in the
example of the non-generic sphere above.
Thus the sum of the asymptotic contributions $I_{\l\m\n}$ collapses
to:
\eqn\clps{\sum_{e} \sum_{(i,j) \in {\l}_e} \left[ m_{1e} (i-1) + m_{2e} (j-1) + 1
\right]}
As we said above the fluctuation determinants cancel
up to a well-defined sign, which can be determined by slightly involved calculation:
\eqn\cngs{(-1)^{[{\CI}]}, \quad [{\CI}] = \sum_{f} |{\pi}_f| + \sum_{e} \sum_{(i,j) \in {\l}_{e}}
\left( m_{1e}(i-1) + m_{2e}(j-1) + 1 \right) + m_{1e} |{\l}_{e}| \ {\rm mod} \ 2}
(recall that $m_1 + m_2 = 2$ so mod $2$ it does not matter which one of $m$'s use). So
all we are left with is the instanton weight \iw.
The vertex factor
 simply reduces to summation over all ${\pi}_f$ with the
asymptotics $\l\m\n$ with the weight
$$
(-1)^I e^{i{\vartheta}I}=(-q)^I
$$
where
\eqn\obyem{I  = \sum_f |{\pi}_f| + \sum_{e} \sum_{(i,j)
\in {\l}_e} \left[ m_{1e} (i-1) + m_{2e} (j-1) +1
\right]}
Combining this with the minus sign on the edges, as in \cngs, we conclude:
\eqn\prtnfnc{Z_{X}(q, t) = \sum_{\{ {\pi}_f \} } (-q)^{I} \prod_{e} (-1)^{m_{e}
|{\l}_{e}|} e^{-t_{e}|\lambda_{e}|}}
which reproduces the partition function for crystal melting
with fixed asymptotes, including the framing factors (the edge signs), provided we identify:
\eqn\idnt{e^{-g_{s}} = - e^{i{\vartheta}}\ .}
\bigskip\bigskip
\boxit{\medskip
\ndt Thus, we have established the complete equivalence of the topological
vertex counting of the all-genus string partition function, with the ``U(1)
maximally supersymmetric topologically twisted gauge theory'' partition
function on toric Calabi-Yau manifolds.}

\newsec{Torsion free sheaves and general Calabi-Yau's}

So, on ${\bC}^3$ we have a well-defined noncommutative gauge theory problem,
and its solution can be phrased in purely commutative terms (for the
holomorphic functions form a commutative subalgebra), i.e. in terms of
ideals. What will replace this structure on a
general manifold $X$, not necessarily
toric? The global object that corresponds locally to an ideal is
called an {\it ideal sheaf}. In each coordinate patch it is given
by an ideal ${\CI}_{U_{\a}}$ in the algebra of holomorphic functions
${\CO}_{U_{\a}}$ on the patch, subject to the obvious
consistency relations. It can be also characterized abstractly
as rank one torsion free sheafs with trivial $c_1$.
Here $c_1$ is the first Chern class of the ideal sheaf
defined using any locally free resolution. Similarly,
one defines $c_2$ and $c_3$, which need not be trivial.

Any ideal sheaf $\CI$ can be blown up. Blowup is a local
construction and locally it can be described as follows.
Suppose the ideal $I$ is generated by the polynomials
$f_{1}(z), \ldots, f_{k}(z)$. Then the blowup ${\hat X}$ of $I$
is the closure in $X \times {\bP}^{k-1}$ of the graph:
\eqn\grh{{\widehat X} = {\overline{\{ (z, (f_1(z) : \ldots : f_{k}(z))) \vert
z \in X \backslash Z \} }}}
where $Z$ is the set of common zeroes of $f_1, \ldots, f_k$. For toric
ideals, the generators $f_1, \ldots, f_k$ are in one-to-one correspondence
with the inner corners of the 3d Young diagram, and the resulting space
${\hat X}$ has as its toric base ${\Delta}({\hat X})$ the convex hull of the
set of these corners. Observe that, by construction, the blowup
comes with an embedding into a projective space and, hence, with
a natural line bundle, K\"ahler metric, etc.

There is a natural moduli space of ideal sheaves ${\CM}^{sheaf}$
on $X$, which is nothing but the Hilbert scheme of curves in $X$.
This space is highly singular and the naive integration of the
top Chern class of the obstruction bundle over it is ill-defined.
However, the philosophy of virtual fundamental classes is
applicable and yields a well-defined canonical zero-dimensional
homology cycle \thom \mnop.

In the case of toric $X$ one can apply equivariant localization
applied to a suitable class to verify
that the sheaf cohomology produce the same result
as the topological vertex \mnop,
in accord with the gauge theory construction described in the previous
section.
This involves the identification of the ``Donaldson-Thomas class''
\thom\
as the relevant object to integrate over the moduli
of ideal sheaves.  It would be very interesting
to relate more explicitly this class with that predicted by our gauge theory.

One can also extend the theory to more general $X$, first of all to compact
Calabi-Yau's where there is no, up to date, effective technique of
calculating the integrals over the virtual
fundamental cycle of ${\CM}^{sheaf}$, except for the trivial $ch_2$ (maps to
the point).

The gauge theory construction, however, provides hope for
computational
progress in
the case of compact Calabi-Yau's, say quintic.
The ground for this hope is the
analogy with the four dimensional case, notably the Seiberg-Witten solution
of the ${\CN}=2$ theory \witsei, more precisely
its recent formulation in terms of
equivariant intersection theory on the moduli space of instantons on
${\bR}^4$ \neksw. This is the analogue of our vertex. So we have now, at our
disposal, an analogue of Seiberg-Witten ansatz (together with all
gravitational couplings, necessary for putting the theory on a curved
background).
What is lacking, if anything, is the analogue of the full
Seiberg-Witten geometry, including the monopole equations.
This would be a fascinating
subject to further develop.

To support this analogy further, and explain what sort of objects we expect
to see in our dual theory, let us recall the behavior of the maximally
supersymmetric Yang-Mills theory in four dimensions. When appropriately twisted, this
theory calculates the Euler characteristics of the appropriately completed and compactified instanton moduli spaces
on the spacetime manifold $X$ \nfour. The partition function, as a function of the
complexified theta angle is a quasi-modular form, in accordance with
S-duality \nfour. For example, for $U(1)$ theory on K3, it is given by:
\eqn\nfkt{Z_{K3} (q, t) = {{\vartheta}_{{\Gamma}^{3,19}} (t, {\tau},
{\bar\tau})\over {\eta}^{24}(q)}}
where $t \in H^{2}(K3, {\bR})$, $ q = {\exp} 2\pi i {\tau}$. The
denominator, which comes from the point-like instantons, is the analogue of
the MacMahon function to the power of ${\chi}/2$ which is the expected
result for topological strings on
 Calabi-Yau threefold.

\bigskip
\ndt {\bf Acknowledgements.}
\bigskip
\ndt We would like to thank the first Simons Workshop on
Mathematics and Physics where this work was initiated. We would
also like to thank S. ~Gukov, K.~Hori, G.~Horowitz, S. ~Katz,
H.~Ooguri and A.~Strominger for valuable discussions. NN is
grateful to the High Energy Group at Harvard University and to the
Physics Department of Rockefeller University for hospitality.

The research of AI and CV is supported in part by NSF grant
DMS-0074329.  CV is additionally supported by NSF grant
PHY-9802709. The research of NN is partly supported by {\cyr RFFI} grant 03-02-17554
and
by the grant {\cyr NSh}-1999.2003.2 for scientific schools.
A.~O.\ was partially supported by
DMS-0096246 and fellowships from the Packard foundation.

\listrefs

\end